\documentclass[journal]{aa}

\usepackage{graphicx}
\usepackage{amsmath}
\usepackage{amssymb}
\usepackage{verbatim}
\usepackage{array}
\usepackage{txfonts}
\usepackage{natbib}
\usepackage[switch,pagewise]{lineno}
\usepackage{multirow}
\usepackage{float}
\usepackage{color}
\usepackage{caption}
\usepackage{xspace}  % This handles spaces after commands    
\usepackage[acronym,nomain]{glossaries}  

\bibpunct{(}{)}{;}{a}{}{,}

\newcommand{\celltspace}{\rule{0pt}{2.2ex}}
\newcommand{\cellbspace}{\rule[-1.1ex]{0pt}{0pt}}

\newcommand{\dunit}{\,cm$^{2}$\,s$^{-1}$}

\newcommand{\kms}{\,km\,s$^{-1}$}
\newcommand{\punit}{\,erg\,s$^{-1}$}

\newcommand{\nunit}{\,H\,cm$^{-3}$}

\newcommand{\msol}{\,M$_{\odot}$}

\newcommand{\snrate}{\,SN\,yr$^{-1}$}
\newcommand{\psrrate}{\,PSR\,yr$^{-1}$}

\newcommand{\gev}{\,GeV\xspace}
\newcommand{\tev}{\,TeV\xspace}
\newcommand{\pev}{\,PeV\xspace}
\newcommand{\mug}{\,$\mu$G}
\def\deg{\ensuremath{^\circ}}

\newacronym{cmb}{CMB}{cosmic microwave background}
\newacronym{cta}{CTA}{Cherenkov Telescope Array}
\newacronym{cr}{CR}{cosmic-ray}
\newacronym{crs}{CRs}{cosmic rays}
\newacronym{gc}{GC}{Galactic center}
\newacronym{gps}{GPS}{Galactic plane survey}
\newacronym{he}{HE}{high-energy}
\newacronym{hgps}{HGPS}{H.E.S.S. Galactic plane survey}
\newacronym{iact}{IACT}{imaging atmospheric Cherenkov telescope}
\newacronym{ic}{IC}{Inverse Compton}
\newacronym{ism}{ISM}{interstellar medium}
\newacronym{isrf}{ISRF}{interstellar radiation field}
\newacronym{lmc}{LMC}{Large Magellanic Cloud}
\newacronym{mw}{MW}{Milky Way}
\newacronym{pwn}{PWN}{pulsar wind nebula}
\newacronym{pwne}{PWNe}{pulsar wind nebulae}
\newacronym{sdr}{SDR}{suppressed diffusion region}
\newacronym{sn}{SN}{supernova}
\newacronym{sne}{SNe}{supernovae}
\newacronym{ccsne}{ccSNe}{core-collapse supernovae}
\newacronym{snr}{SNR}{supernova remnant}
\newacronym{snrs}{SNRs}{supernova remnants}
\newacronym{uhe}{UHE}{ultra-high-energy}
\newacronym{vhe}{VHE}{very-high-energy}

\begin{document}

% Title, header, abstract
\title{Population synthesis of pulsar wind nebulae\\and pulsar halos in the Milky Way}
\subtitle{Predicted contributions to the very-high-energy sky}
\author{Pierrick Martin\inst{\ref{irap}} 
\and Luigi Tibaldo\inst{\ref{irap}}
\and Alexandre Marcowith\inst{\ref{lupm}}
\and Soheila Abdollahi\inst{\ref{irap}}}
\authorrunning{Pierrick Martin et al.}
\institute{
IRAP, Universit\'e de Toulouse, CNRS, CNES, F-31028 Toulouse, France \label{irap}
\and Laboratoire Univers et Particules de Montpellier (LUPM) Universit\'e Montpellier, CNRS/IN2P3, CC72, \\Place Eug\`ene Bataillon, F-34095 Montpellier Cedex 5, France \label{lupm}
}

\date{Received 11 May 2022 / Accepted 06 July 2022}
\abstract{The discovery of extended gamma-ray emission toward a number of middle-aged pulsars suggests the possibility of long-lived particle confinement beyond the classical pulsar wind nebula (PWN) stage. How this emerging source class can be extrapolated to a Galactic population remains unclear.}
{We aim to evaluate how pulsar halos fit in existing TeV observations, under the assumption that all middle-aged pulsars develop halos similar to those observed toward the J0633+1746 or B0656+14 pulsars.}
{We modeled the populations of supernova remnants, PWNe, and pulsar halos in the Milky Way. The PWN-halo evolutionary sequence is described in a simple yet coherent framework, and both kinds of objects are assumed to share the same particle injection properties. We then assessed the contribution of the different source classes to the very-high-energy emission from the Galaxy.}
{The synthetic population can be made consistent with the flux distribution of all known objects, including unidentified objects, for a reasonable set of parameters. The fraction of the populations predicted to be detectable in surveys of the Galactic plane with H.E.S.S. and HAWC is then found to be in good agreement with their actual outcome, with a number of detectable halos ranging from 30 to 80\% of the number of detectable PWNe. Prospects for CTA involve the detection of $250-300$ sources in the Galactic Plane Survey, including 170 PWNe and up to 100 halos. The extent of diffusion suppression in halos has a limited impact on such prospects but its magnitude has a strong influence. The level of diffuse emission from unresolved populations in each survey is found to be dominated by halos and comparable to large-scale interstellar radiation powered by cosmic rays above 0.1--1\tev.}
{Pulsar halos are shown to be viable counterparts to a fraction of the currently unidentified sources if they develop around most middle-aged pulsars. Yet, if the phenomenon is rare, with an occurrence rate of $5-10$\% as suggested in a previous work from the local positron flux constraint, the total number of currently known TeV sources including unidentified ones cannot be accounted for in our model from young PWNe only. This calls for continued efforts to model pulsar-powered emission along the full evolutionary path, including the late stages past the young nebula phase.}
\maketitle

% Introduction
\section{Introduction}
\label{intro}

Over the past decade, it has become clear that pulsars are major players in the \gls{vhe} sky, and possibly even in the \gls{uhe} sky \citep{Abdalla:2018a,Albert:2021,Breuhaus:2022,DeOnaWilhelmi:2022}. While formal identification is still lacking for a large fraction of the constantly increasing population of known \gls{vhe} and \gls{uhe} gamma-ray sources, observational evidence is growing in support to this conclusion.

Emission components detected in the \gls{hgps} are found to be significantly correlated in position with energetic pulsars, while this correlation is absent for less energetic pulsars \citep{Abdalla:2018b}. Out of the 78 sources detected in the \gls{hgps}, 42 can be positionally associated with an energetic pulsar \citep{Abdalla:2018a}; among those, 14 are clearly identified as \gls{pwne} produced by young and energetic pulsars, with characteristic ages of $\lesssim 50-100$\,kyr, and an additional ten are solid candidates \citep{Abdalla:2018b}. At slightly higher energies and over a shifted portion of the plane, 15 out of 39 sources listed in the second catalog of HAWC sources \citep[2HWC;][]{Abeysekara:2017b} are spatially coincident with a pulsar \citep{Linden:2017}, and an additional 14 positional associations are found among the 20 new sources with no previous TeV counterpart listed in the third catalog of HAWC sources \citep[3HWC;][]{Albert:2020}. At even higher energies, ten out of the 12 sources detected with LHAASO in the $\sim0.1-1$\pev range are positionally coincident with energetic pulsars \citep{Cao:2021}. 

In all cases, the association criteria are rather loose and cannot prevent chance coincidence or erroneous associations, in part due to the significant extension of most Galactic sources at these energies. On the other hand, the sample of known pulsars used in the correlations mostly consists of objects identified from their beamed emission and, for each $\sim10-100$\,kyr pulsar with a radio or X-ray beam crossing our line of sight, there are three to four misaligned pulsars that would go undetected in these bands and they could give rise to unbeamed gamma-ray emission \citep{Linden:2017}. 

Although the exact chain of processes by which this happens has not been completely revealed, pulsars seem to be able to convert a significant fraction of their rotational energy into the acceleration of electron-positron pairs with energies reaching 100\tev and above \citep{Kargaltsev:2015,Amato:2020}. Electron and positrons pairs accelerated by pulsar systems are eventually released into the interstellar medium after some confinement close to the source. The conditions of this confinement -- such as spatial extent, time evolution, and energy-dependent nature -- have important observable consequences. They determine how these leptons will contribute to the overall nonthermal emission of the Galaxy, in spectral and angular distribution, and, for pulsars in our vicinity, they drive the amount and spectrum of energetic particles that can be detected at Earth, which has implications on the search for dark matter by-products in the local flux of cosmic rays \citep{Hooper:2017,Profumo:2018,Manconi:2020}.

Accelerated pairs are efficiently trapped in a hot and highly magnetized shocked pulsar wind bubble over the first few $1-10$\,kyr \citep{Gaensler:2006}. These \gls{pwne} are the most likely origin for a (possibly large) fraction of the sources detected in the Galactic plane, for instance with H.E.S.S. \citep{Abdalla:2018b,Abdalla:2018a}. The recent discovery of extended emission around two nearby middle-aged pulsars, with characteristic ages $\gtrsim 100$\,kyr, suggests the possibility of further confinement beyond the nebula stage \citep{Abeysekara:2017b}. Since then, a number of other candidates were suggested for these so-called TeV-halos or gamma-ray halos \citep{DiMauro:2020,Albert:2020,Aharonian:2021}, and the phenomenon was demonstrated to have a broadband signature, at least in the gamma-ray range from below 10\gev up to above 100\tev \citep{DiMauro:2019a,Aharonian:2021}. Theoretically, the question remains essentially open as to how exactly efficient confinement in the vicinity of the pulsar is achieved \citep{Evoli:2018,LopezCoto:2018,Giacinti:2020,Fang:2019,Mukhopadhyay:2022} -- in which medium, by which physical mechanism, and over which extent and duration -- and how the most solid examples of this emerging source class could be extrapolated to a Galactic population of objects located in a variety of environments. 

Even if the physics driving gamma-ray halos still remains to be elucidated for the most part, the source class holds promise for a better and more complete understanding of the \gls{vhe} and \gls{uhe} sky. If most middle-aged pulsars in the Galaxy happen to develop spatially extended and long-lived halos (still a bold assumption at that stage), the total contribution could end up being non-negligible. The brightest halos could already be present as unidentified or unassociated sources in the gamma-ray source catalogs; their identification, however, could turn out to be quite tricky owing to the complex morphology that can result from the combined effects of proper motion, injection history, and actual structure of the surrounding medium \citep{Zhang:2021}. For those halos below the current detection thresholds, they could account as unresolved sources for a fair fraction of the diffuse emission detected in regions of the Galactic plane \citep{Linden:2018} or toward the Galactic center \citep{Hooper:2018,Hooper:2022}.

In this paper, we aim to provide a quantitative assessment of how pulsar halos fit in the Galactic population of \gls{vhe} sources, especially in relation to the \gls{pwne} population. Specifically, we want to evaluate: (i) whether it is conceivable that most middle-aged pulsars in the Galaxy develop halos similar to the few instances known today; (ii) the fraction of the currently unidentified or unassociated \gls{vhe} sources that could actually be halos; (iii) how many halos could become detectable in forthcoming surveys of the Galactic plane; (iv) how the cumulated emission of unresolved halos compares to that resulting from \gls{crs} interacting with the \gls{ism}, and to that of other unresolved source classes.

We start by introducing in Sect. \ref{model} the frameworks used to model individual objects, and then present in Sect. \ref{popmod} how these were combined to generate a mock population for the whole Milky Way. We discuss in Sect. \label{popres} the properties of one statistical realization of the synthetic population and assess the prospects for observing them, either as individually resolved objects or as an unresolved diffuse emission. Last, we summarize the main findings of our study in Sect. \ref{conclu}.

% Command for merging two cells horizontally
% \celltspace Diffusion rigidity scaling index $\delta_{\textrm{D}}$ &\multicolumn{2}{c |}{$1/3$} \\
\begin{table*}[t!]
\centering
\begin{tabular}{| c | c |}
\hline
\celltspace Parameter & Value \cellbspace \\
\hline
\multicolumn{2}{| c |}{\celltspace PWNe \cellbspace} \\
\hline
\celltspace Nebular magnetic field initial strength $B_0$ (G) & $5 \times 10^{-5}$ \\
\celltspace Nebular magnetic field evolution index $\delta_{\textrm{B}}$ & 0.6 \\
\celltspace Injection distribution $S_{\rm PWN}$ & BPLEC \\
\celltspace Injection distribution index below break $\alpha_1$ & 1.5 \\
\celltspace Injection distribution index above break $\alpha_2$ & $\mathcal{U}(2.4,0.4)$ \\
\celltspace Injection distribution break energy $E_{\rm brk}$ (GeV) & $100$ \\
\celltspace Injection distribution cutoff energy $E_{\rm cut}$ (TeV) & $\mathcal{U}(500,300)$  \\
\celltspace Injection efficiency $\eta_{\textrm{PWN}}$ (\%) & $\mathcal{U}(70,30)$ \\
\celltspace PWN age limit $\tau_{\textrm{PWN}}$ (yr) & $10^5$ \cellbspace\\
\hline
\multicolumn{2}{| c |}{\celltspace Halos \cellbspace} \\
\hline
\celltspace Suppressed diffusion region size $R_{\textrm{SDR}}$ (pc)  & 30 or 50 or 80 \\
\celltspace Suppressed diffusion normalization at 100\tev $D_0^{\textrm{SDR}}$ (\dunit) & $4 \times 10^{27}$ or $4 \times 10^{28}$ \\
\celltspace Average interstellar diffusion normalization at 100\tev $D_0^{\textrm{ISM}}$ (\dunit) & $2 \times10^{30}$ \\
\celltspace Diffusion rigidity scaling index $\delta_{\textrm{D}}$ & $1/3$ \\
\celltspace Injection distribution $S_{\textrm{HALO}}$ & $S_{\textrm{HALO}} = S_{\textrm{PWN}}$ \\
\celltspace Injection efficiency $\eta_{\textrm{HALO}}$ & $\eta_{\textrm{HALO}} = \eta_{\textrm{PWN}}$ \\
\celltspace Injection start time $\tau_{\textrm{INJ}}$ (yr) & $\tau_{\textrm{INJ}} = \tau_{\textrm{EXIT}}$ \\
\celltspace Halo age limit $\tau_{\textrm{HALO}}$ (yr) & $4 \times10^{5}$ \cellbspace\\
\hline
\multicolumn{2}{| c |}{\celltspace SNRs \cellbspace} \\
\hline
\celltspace Ejecta mass $M_{\textrm{ej}}$ (\msol) & 1.4\msol/5\msol\ for SNe Ia/ccSNe \\
\celltspace Ejecta energy $E_{\textrm{ej}}$ (erg) & $\mathcal{L}(51,0.5)$ \\
\celltspace Injection distribution $S_{\textrm{SNR}}$ & PLEC \\
\celltspace Injection efficiency $\eta_{\textrm{SNR}}$ (\%) & $\mathcal{U}(10,30)$ \\
\celltspace Injection index $\alpha_{\rm e,p}$ & $\mathcal{U}(2.2,2.4)$ \\
\celltspace Injection electron-to-proton ratio $K_{\rm ep}$ & $10^{-3}$ \\
\celltspace SNR age limit $\tau_{\textrm{SNR}}$ (yr) & $3 \times 10^4$ \cellbspace\\
\hline
\end{tabular}
\caption{Summary of parameters used in the modeling of \gls{pwne}, halos, and SNRs}
$\mathcal{U}(\mu,\sigma)$ indicates a uniform distribution with mean $\mu$ and standard deviation $\sigma$.\\
$\mathcal{L}(\mu,\sigma)$ indicates a log-normal distribution with mean of the logarithm $\mu$ and standard deviation of the logarithm $\sigma$.\\
(B)PLEC stands for (broken) power law with exponential cutoff 
\label{tab:modpars} 
\end{table*}

% Single SNR-PWN-halo models
\section{Single SNR-PWN-halo models}
\label{model}

We describe in this section the main assumptions and parameters underlying our models for an individual \gls{pwn}, pulsar halo, or \gls{snr}.

% Pulsar wind nebulae
\subsection{Pulsar wind nebulae}
\label{model:pwn}

The modeling of the \gls{pwne} population is based on the individual \gls{pwn} model introduced in \citet{Mayer:2012} and subsequently corrected in \citet{Abdalla:2018b}. Appendix A in the latter reference provides a complete summary of the formalism subtending the model and here we just refer to the main equations of the model.

The modeling starts from essential pulsar properties (synthetic or inferred from observations), such as spin period, magnetic field, and braking index, that determine the spin-down history of the compact object, hence the time evolution of the maximum power available for nonthermal particle injection into the nebula or halo. Following Eqs. 3 and 4 in \citet{Martin:2022}, the latter is assumed to occur with constant efficiency over the pulsar's lifetime, and to have a constant broken power-law spectral shape for each pulsar, with a harder distribution below a break energy of 100\gev, and a softer one above it \citep{Zhang:2008,Bucciantini:2011,Torres:2014}. The minimum particle energy is set at 1\gev, while the maximum is determined by an exponential cutoff randomly selected in a uniform distribution extending from 200 to 800\tev. Such a prescription results in cutoff energies that are on average a factor $5-6$ below the maximum possible particle energy under the assumption of ideal magnetohydrodynamics for spin-down powers in the $10^{36}-10^{37}$\punit range \citep{DeOnaWilhelmi:2022}. In the range $10^{34}-10^{36}$\punit most relevant for pulsar halos, our randomly sampled cutoff energies are on average within $\pm 50$\% of this physical limit. At low spin-down powers $10^{33}-10^{34}$\punit , the cutoff energies generally exceed the limit, by a factor 4 on average, but this range contributes little to the \gls{vhe} landscape. Increasing measurements at ultra high energies with LHAASO will be instrumental in characterizing the maximum particle energies attained by pulsars in different regimes \citep{Cao:2021}.

The \gls{pwn} model prescribes the expansion of a spherical nebula over three consecutive dynamical stages: (i) rapid supersonic expansion in the unshocked supernova ejecta, powered by nearly constant spin-down, until the latter starts declining; (ii) constant-speed expansion, as the nebula receives little energy from the pulsar, until the reverse shock from the remnant crushes the nebula; (iii) subsonic expansion into the hot shocked ejecta \citep[an alternative sequence occurs in the rarer cases where the reverse shock crushing takes place before the spin-down time scale; see Eqs A.12 and A.13 in][]{Abdalla:2018b}. Quantitatively, the size evolution of the \gls{pwn} is set by the radius of the nebula at the pulsar spin-down time scale and the reverse shock interaction time scale, both of which are determined from Eqs. 22 and 29 presented in \citet{Reynolds:1984}. 

This model neglects the reverse shock crushing of the nebula and its subsequent reverberation, an evolutionary phase that can bring in some complexity to the dynamical and radiative evolution of the \gls{pwn}, even in the simple case of spherical symmetry without the effects of pulsar motion and asymetric remnant evolution \citep{Bucciantini:2003}. A \gls{pwne} population synthesis including reverberation in the dynamical evolution of the nebula, and a thorough discussion on the current knowledge and uncertainties on this evolutionary phase, can be found in \citet{Fiori:2022}.

The prescribed dynamics of the spherical nebula is used to compute the evolution of its nonthermal particles content. The latter results from the combined time-dependent injection by the pulsar, synchrotron, inverse-Compton and adiabatic energy losses, and diffusive escape from the nebula. For synchrotron radiation, the model includes a flux-conserving time evolution for the uniform nebular magnetic field, starting from an initial value treated as free parameter \citep[see Eq A.14 in][]{Abdalla:2018b}. 

In our joint model for a pulsar nebula and halo (to be introduced in more details below), we assume that particle injection in the nebula ends when the pulsar exits it as a result of its kick velocity. This marks the beginning of the pulsar halo phase, while the nebula continues its evolution as a relic structure, not fed anymore with freshly accelerated particles (and not powered anymore by the pulsar's spin-down, although this is less important as exit from the nebula usually occurs at times $\gtrsim10-20$\,kyr, when the pulsar has gone relatively weak). The evolution of the \gls{pwn} is computed until a maximum age $\tau_{\textrm{PWN}}$ = 100\,kyr. 

The main model parameters are summarized in Table \ref{tab:modpars}. These include a number of free parameters that were set by comparison of the population synthesis predictions with the properties of the sample of objects observed at very high energies, starting from initial guesses informed by studies of individual objects \citep[e.g.,][]{Zhang:2008,Bucciantini:2011,Mayer:2012,Torres:2014}.

% Pulsar halos
\subsection{Pulsar halos}
\label{model:halo}

The detection with MILAGRO, HAWC, and now LHAASO of extended gamma-ray emission structures around some middle-aged pulsars \citep{Abdo:2009,Abeysekara:2017b,Aharonian:2021}, subsequently dubbed TeV halos or gamma-ray halos \citep{Linden:2017,Linden:2018}, has generated some debate about the exact physical setup at stake in these objects and its relation to the evolution of pulsar-\gls{pwn}-\gls{snr} systems. A review of the phenomenon and its potential as a new and distinct source class in high-energy astrophysics can be found in \citet{LopezCoto:2022}.

An important question is that of the physical state of the medium in which the pair halo develops, which is particularly relevant to any attempt to account for the efficient confinement of particles in the vicinity of the pulsar. The various possibilities envisioned so far include: (i) the undisturbed \gls{ism}, where standard magnetohydrodynamical turbulence happens to have the required properties for efficient scattering of $\sim100$\tev particles, for instance a small pc-scale turbulence correlation length \citep{LopezCoto:2018,Giacinti:2020}; (ii) the relic nebula, the specific magnetic topology and relatively high fields of which would cause the trapping of particles \citep{Tang:2019}; (iii) the parent \gls{snr}, the expansion and evolution of which is a source of fluid turbulence downstream of the forward shock, eventually responsible for the suppressed diffusion close to a pulsar that did not escape its parent remnant \citep{Fang:2019}; (iv) a stellar wind-blown bubble, the interior of which features a high level of standard magnetohydrodynamical turbulence \citep{Fang:2019}.

The above scenarios are all likely to be realized in nature, but they may have different consequences in terms of halo properties and evolution and extrapolation to a galactic population. It is beyond the scope of this work to explore all these alternatives and their combinations, and we instead adopted a physical setup for a halo in the spirit of the picture sketched in \citet{Giacinti:2020}: (i) the halo phase starts when the pulsar becomes supersonic in its surrounding medium and enters the bow-shock phase, which could happen either within the remnant after leaving the relic nebula or out of it when crossing the forward shock; (ii) relativistic pairs accelerated in the pulsar and its wind nebula can easily escape, almost unaffected by energy losses, and are injected  isotropically in the surrounding medium; (iii) particles are free to propagate diffusively in the surrounding medium, either the remnant interior or the \gls{ism} essentially undisturbed, except for the possibility of a suppressed diffusion of unspecified origin.

The full model provides some coherence at the population level for the \gls{pwne} and halos classes, in particular because they share a common prescription for the injection and follow a physically motivated evolutionary sequence. Yet, it has a limited capability to capture the complexity of some individual objects, especially the transitional ones moving from nebula to halo after dispersion of the former by the reverse-shock crushing. 

We note that the delayed injection scheme has two important consequences: (i) it reduces the impact on the results of uncertainties in the pulsar's spin-down history, especially at early times when most of the rotational energy is lost ; (ii) the morphology of the halo can be expected to be more compact and may display less complicated patterns than those considered for instance in \citet{Zhang:2021}.

In practice, the modeling of individual halos is based on the phenomenological two-zone model presented in \citet{Tang:2019} and \citet{DiMauro:2019a} in the context of Geminga. We briefly describe here the main points of the model and refer the reader to these articles for the full formalism. The main equations of the model and a thorough discussion of the parameters adopted can be found in \citet{Martin:2022}.

Particle injection from a subparsec bow-shock nebula is treated as point-like in space and the effect of proper motion over the lifetime of the halo is not included because we are not interested here in detailed morphological aspects \citep[that can be quite complex; see][]{Zhang:2021}. Injection starts when the pulsar develops a bow-shock, typically $40-60$\,kyr after birth \citep{Bykov:2017,Evoli:2021}, and it proceeds with a power being a constant fraction of the declining spin-down luminosity of the pulsar, with typical values on the order of a few tens of percent \citep{Bucciantini:2011,Torres:2014}. The injection spectrum is assumed to be a broken power law, similar in shape and acceleration efficiency to that inferred for \gls{pwne}. Considering that high injection efficiencies close to 100\% inferred for kyr-old objects could still hold for the 100\,kyr-old pulsars powering halos may appear as a bold extrapolation. Yet, as demonstrated in \citet{Martin:2022}, the injection efficiencies required for the two canonical halos around J0633+1746 and B0656+14 are of that order.

Particles diffuse away spherically in a medium characterized by a two-zone concentric structure for diffusion properties, with an outer region representative of the large-scale average \gls{ism} in the Galactic plane, and an inner region with a radial extent of a few tens of pc where diffusion is suppressed. For both the inner and outer regions, we assumed a diffusion coefficient with power-law rigidity dependence with index $1/3$, applicable for scattering in magnetic turbulence with a Kolmogorov spectrum. The normalization of the diffusion coefficient in the average \gls{ism} is set to $2 \times 10^{30}$\dunit at 100\tev, in agreement with the values obtained in fits of diffusion+reacceleration propagation scenarios to a large set of observables \citep{Trotta:2011,Orlando:2018}. 

Along their propagation, particles lose energy and radiate via the synchrotron and inverse-Compton scattering processes in interstellar magnetic and radiation fields, the distribution of which across the Galaxy will be described below. Inverse-Compton scattering emission is computed using the Naima python package \citep{Zabalza:2015}, in the isotropic seed photon field approximation \citep[see][for an example of anisotropic calculation in the case of the Geminga halo]{Johannesson:2019}.

Injection parameters are mostly set by past studies of \gls{pwne}, and environmental parameters are mostly defined by large-scale models of the Galactic \gls{ism}. Eventually, the parameters most specific to halos are the extent of the suppressed diffusion region and the diffusion coefficient therein. Possible values for these were determined from gamma-ray observations of the halos around the J0633+1746 and B0656+14 pulsars. In \citet{Martin:2022}, various halo model setups for either object were jointly fit to the angular intensity profile inferred from HAWC \citep{Abeysekara:2017b} and the integrated spectrum obtained with the {\it Fermi}-LAT \citep{DiMauro:2019a}. An additional constraint in this exercise was to make sure that the escaping positron flux from either halo does not exceed the local positron flux measured with AMS-02 \citep{Aguilar:2014,Aguilar:2019a}. From this work, we retain as possible scenarios for diffusive halos: three possible sizes (30, 50 and 80\,pc), and two levels of diffusion suppression (by a factor 500, as indicated by Geminga, and by a factor 50, in agreement with B0656+14 within uncertainties). 

Table \ref{tab:modpars} summarizes the full list of parameters characterizing these scenarios in our halo population synthesis. By default, we assume that all middle-aged pulsars in the bow-shock phase develop a halo. We discuss below the implications of an occurrence rate for pulsar halos much smaller than 100\%, as was suggested in \citet{Martin:2022} from the realization that the local positron flux produced in a scenario where all nearby pulsars develop a halo may exceed the AMS-02 measurement. We compute the evolution of halos until a maximum age $\tau_{\textrm{HALO}}$ = 400\,kyr. Depending on the actual processes driving the existence of pulsar halos, a minimum spin-down power could be a more relevant stopping criterion than a maximum age but our model setup does not link halo properties to pulsar properties. Adopting a 400\,kyr age limit encompasses all known canonical halos and include in the population those objects providing a significant contribution to the detectability prospects, as illustrated in Fig. \ref{fig:res:lum}, while keeping computation time within reasonable limits.

% Supernova remnants
\subsection{Supernova remnants}
\label{model:snrs}

Although not the main focus of this work, our population model also includes a component for \gls{snrs} that we briefly describe here, in particular because the evolution of the parent \gls{snr} determines the dynamics of the \gls{pwn} it contains (for pulsar-producing core-collapse supernovae). The modeling of \gls{snrs} is based on the individual \gls{snr} model presented in \citet{Cristofari:2013,Cristofari:2017}, and the full formalism can be found in that reference. We just recall here the working of the model and introduce the values considered for its main parameters. 

The model implements analytical prescriptions for the dynamics of the forward shock in the remnant \citep[Eqs. 1--4 in][]{Cristofari:2013}. Different treatments are used depending on whether the \gls{snr} results from a thermonuclear or core-collapse explosion (hereafter SNe Ia or ccSNe): in the former case, the expansion occurs in a uniform circumstellar medium, while in the latter case it occurs in a layered wind-blown cavity shaped by the progenitor massive star. 

Remnants of each type are assumed to have the same ejecta mass, 1.4\msol\ for SNe Ia and 5\msol\ for ccSNe, but their ejecta kinetic energies are sampled from a log-normal distribution with mean $\mu_{log(E_0)}=51$ and standard deviation $\sigma_{log(E_0)}=0.5$ for $E_0$ the supernova explosion energy in erg. Such a distribution is close to that inferred from the properties of X-ray \gls{snrs} in the Large Magellanic Cloud and the Milky Way \citep{Leahy:2017,Leahy:2020}. For remnants of ccSNe, the same progenitor stellar wind properties were assumed \cite[see the parameters in][]{Cristofari:2013}, which is most likely not realistic, but variety of the dynamics over the population ensues nevertheless from the random-selected ejecta energy and outer interstellar density of each mock \gls{snr} (see below).

From the prescribed dynamics, the model defines a parameterized distribution of accelerated particles at the shock at each time step, under the assumptions that it has a power-law shape with exponential cutoff and fixed index and amounts to a given level of pressure at the shock. Accelerated particles are assumed to follow a power-law in energy starting from 1\,GeV and extending up to the maximum allowed energy that results from upstream escape in the case of protons, and from energy losses in the case of electrons \citep[Eqs. 7 and 15 in][]{Cristofari:2013}. In the case of electrons, a cooling break can appear at an energy at which the loss time scale is smaller than the remnant's age \citep[Eq. 16 in][]{Cristofari:2013}.The power-law index (below the break energy if there is one in the electron spectrum)  is a free parameter and, in our population model, it was randomly selected for each \gls{snr} from a uniform distribution between 2.2 and 2.4. As acceleration efficiency parameter, we used the definition of Eq. 13 in \citet{Ellison:2000}, neglecting escape flux and upstream pressure (the latter point may need critical examination in the context of remnants from ccSNe, which expand in hot wind-blown bubbles).

When advancing in time, particles at the shock are advected downstream and a whole population of accelerated particles progressively builds up in concentric layers, filling the remnant from the center out to the forward shock, and contributing to the nonthermal emission of the remnant: inverse-Compton emission is computed in the radiation fields assumed to bath the system, while pion-decay emission is computed in the assumed radial density distribution of the remnant \citep[Eq. 6 in][combined with the continuity equation]{Cristofari:2013}. The model is valid over the free expansion and Sedov-Taylor stages and breaks down as the forward shock becomes radiative. Adiabatic losses for spherical expansion is included for those particles trapped within the remnant and accompanying its expansion. We assumed a lifetime of $\tau_{\rm SNR}=30$\,kyr for model \gls{snrs} but some do not even reach that limit as they become radiative before.

When calibrating our population model from Galactic observations of \gls{vhe} objects (see below), acceleration efficiencies extending to relatively high values were required: we found that a uniform distribution of acceleration efficiencies between 0.1 and 0.3 and a fixed electron-to-proton number ratio of $10^{-3}$ provided an acceptable match to the observed flux distribution for Galactic \gls{snrs} in the TeV range. Several caveats should be mentioned though: (i) this does not come out of a formal fit over the full parameter space of the \gls{snrs} population, so it may well be that a range of lower values provides a good match; (ii) in particular, the previous point depends on the assumed spatial distribution for SNRs, and especially the description of our local environment, which is a shortcoming of the adopted spatial distribution model; (iii) the model does not describe remnants interacting with dense gas clouds, which are a non-negligible fraction of the observed population \citep{Abdalla:2018a}; (iv) the $0.1-0.3$ efficiencies apply to the full range of particle energies, $\geq1$\gev, but the model was not tested against observations in lower-energy bands such as GeV, X-rays, or radio.

% Galactic population model
\section{Galactic population model}
\label{popmod}

In this section, we introduce the assumptions subtending the construction of a mock population of \gls{pwne} and pulsar halos from the above model for individual objects.

% Young pulsar population
\subsection{Young pulsar population}
\label{popmod:psr}

The modeling of the Galactic population of \gls{pwne} and halos starts with the random generation of a pulsar population. We followed an approach similar to that of \citet{Sudoh:2019} for pulsar halos, or \citet{Johnston:2020} for radio and GeV young pulsars.

The source population model starts with the random generation of \gls{ccsne} over the last 400\,kyr (the lifetime of the longest-lived objects in our population synthesis, pulsar halos): random generation of a number of \gls{sne} events from a Poisson distribution with mean $r_{\textrm{SN}} \times \tau_{\textrm{HALO}}$, random generation of an age in a uniform distribution, then random generation of a \gls{sn} type in a binomial distribution and finally, for \gls{ccsne}, random selection of those giving birth to pulsars again from a binomial distribution. We assumed an \gls{sn} rate of $r_{\textrm{SN}}=0.02$\snrate\ \citep{Tammann:1994}, constant over the past 400\,kyr, with a ratio of core-collapse to thermonuclear \gls{sne} of 2 \citep{vandenBergh:1991}, and a pulsar-producing core-collapse SNe rate of $r_{\textrm{PSR}}=0.01$\psrrate \citep{Johnston:2020}. The population of mock pulsars thus obtained is distributed over the Galaxy following the four-spiral-arm pattern used in \citet{Faucher-Giguere:2006}, using the axisymetric radial distribution from \citet{Lorimer:2004} and as vertical distribution that of molecular gas at the solar circle from \citet{Bronfman:1988}.

For each pulsar, an initial pulsar spin period is sampled from a normal distribution with $\mu_{P_0}=50$\,ms and standard deviation $\sigma_{P_0}=35$\,ms, in agreement with the observed population of young and energetic pulsars \citep{Watters:2011,Johnston:2020}. Initial periods are truncated at the centrifugal breakup limit of 0.85\,ms \citep{Sudoh:2019}. Following \citet{Faucher-Giguere:2006} and \citet{Watters:2011}, the initial magnetic field is sampled from a log-normal distribution with $\mu_{log(B_0)}=12.65$ and standard deviation $\sigma_{log(B_0)}=0.55$ with $B_0$ in units of G. The braking index is assumed to be $n=3$, the same for all pulsars, corresponding to a spin-down due to dipole radiation only. Assuming typical values of $10^{45}$\,g\,cm$^2$ and 12\,km for the neutron star inertia and radius, these properties determine the spin-down history of each pulsar, which sets the maximum power available at each time for nonthermal particle injection into \gls{pwne} and halos. All the above parameters are bound to have an impact on the outcome of our population synthesis. A particularly critical aspect, especially for \gls{pwne}, is the description of the spin-down history in the first few kyr of the pulsar. The $n=3$ assumption is not applicable to all young pulsars and all times along their evolution \citep{Parthasarathy:2020}, suggesting a variety of paths for the rotational energy release.

In addition to the above properties, we associate each pulsar to a natal kick. In the context of our model, natal kicks will determine when a pulsar exits its original nebula, which marks the beginning of the halo phase. We used the velocity distribution of young pulsars from \citet{Verbunt:2017}, inferred from very-long-baseline-interferometry parallax measurements of  proper motions. Specifically, we used their two-Maxwellian mixed model for young pulsars with characteristic ages $<10$\,Myr. This distribution has 32\% (68\%) of the pulsars described by a Maxwellian distribution with average velocity 130\kms\ (520\kms).

% Interstellar medium properties
\subsection{Interstellar medium properties}
\label{popmod:ism}

A description of interstellar conditions in the $<50-100$\,pc vicinity of each object is needed for the following reasons: (i) gas density influences the development of parent \gls{snrs}, hence the dynamical evolution of \gls{pwne}; (ii) magnetic and radiation fields set the level of energy losses for nonthermal electrons and positrons; (iii) radiation fields shape the inverse-Compton spectrum. 

As Galactic \gls{isrf}, we adopted the axisymetric model introduced in \citet{Popescu:2017}. In the absence of a robust physical model for the magnetic field on small scales, for instance in the vicinity of star clusters or in superbubbles, we considered that its strength in the close neighborhood of each \gls{pwn}-halo is the large-scale interstellar one. We described the latter as a double exponential model \citep{Strong:2000}, with a radial (vertical) scale length of 6\,kpc (2\,kpc) and a peak value of 12\mug. In practice, this yields magnetic field values of $4-7$\mug\ in the molecular ring, where a large fraction of the pulsars is found, and of 3\mug\ at the solar circle. 

Conversely, interstellar gas density is not described with a complete spatial model over the Galaxy, but as a statistical distribution relevant to the vicinity of parent \gls{snrs}. The latter is inspired by a systematic study of X-ray \gls{snrs} in the Large Magellanic Cloud and Milky Way \citep{Leahy:2017,Leahy:2020} and was assumed to have a log-normal form, with mean $\mu_{log(n_{\rm H})}=0.0$ and standard deviation $\sigma_{log(n_{\rm H})}=0.9$ for $n_H$ in units of \nunit. The mean differs from the value inferred in \citet{Leahy:2017} or \citet{Leahy:2020} and was actually adjusted to have our population of gamma-ray emitting \gls{snrs} match the observed flux distribution (see Sect. \ref{popmod:calib} for a discussion).

% Population calculation
\subsection{Population calculation}
\label{popmod:calc}

The calculation starts with the sampling of SNe events in space and time and the determination of interstellar conditions at their positions. \gls{snrs} are then computed first because the properties of the core-collapse ones sets the dynamics of \gls{pwne} (via the reverse-shock crushing time estimate).

For each mock pulsar of given age $\tau_{\textrm{PSR}}$, we compute the time $\tau_{\textrm{EXIT}}$ at which it escapes the spherical \gls{pwn} that initially developed at the center of the parent \gls{snr}. The dynamics of the latter is determined by the random-selected explosion energy and circumstellar density of the system (ejecta mass being the same for all objects, 5\msol), which in turn sets the dynamics of the former via the reverse-shock crushing time. The calculations performed depend on the relative values of the different ages. If $\tau_{\textrm{PSR}} < \tau_{\textrm{PWN}}$, the evolution of a \gls{pwn} is computed; it can be a relic one if $\tau_{\textrm{PSR}} > \tau_{\textrm{EXIT}}$, or an active one otherwise. If $\tau_{\textrm{EXIT}} < \tau_{\textrm{PSR}} < \tau_{\textrm{HALO}}$: the evolution of a halo is computed, with injection starting at $\tau_{\textrm{EXIT}}$; the halo can coexist with a relic \gls{pwn} if $\tau_{\textrm{PSR}} < \tau_{\textrm{PWN}}$. The full time evolution of each mock \gls{pwn} or halo is computed from the models introduced in Sect. \ref{model:pwn} and \ref{model:halo} until age $\tau_{\textrm{PSR}}$.

With the assumptions exposed so far, a complete steady-state population features on average about 1000 \gls{pwne} and 2600 halos, fewer than expected from the product of pulsar birth rate and halo lifetime because, in a significant number of cases, the pulsar never exits its nebula and so the halo phase never starts. The latter result is the combined effect of a non-negligible fraction of \gls{pwne} growing to relatively large physical sizes, with radii in excess of $30-40$\,pc \citep[in agreement with the limited population known today; see][]{Abdalla:2018b}, and more than 30\% of pulsars having velocities below 200\kms, owing to the low-velocity component in the pulsar velocity distribution from \citet{Verbunt:2017}.

% Population calibration
\subsection{Population calibration}
\label{popmod:calib}

The resulting predicted emission was compared to observations in order to validate or refine the main parameters of the full model, focusing in this exercise on the flux distribution in each source class. The flux distribution of real TeV objects was inferred from existing observations in parts of the Galactic plane \citep[with most of the population statistics being contributed by the \gls{hgps}; see][]{Abdalla:2018a}, and compiled in a modified version of the gamma-cat catalog\footnote{\url{https://gamma-cat.readthedocs.io}}, developed for prospect studies of the future Galactic Plane Survey with \gls{cta} \citep{Remy:2022}. The results for the final set of parameters will be extensively reviewed in the next sections, and we just discuss here those parameters that needed some tuning for the population synthesis to yield a satisfactory description of observations.

We first emphasize that the model optimization does not result from a formal fit over the full parameter space, which would have been prohibitive in terms of computing time. Instead, we proceeded by educated guess, starting from typical parameters for \gls{pwne} and \gls{snrs}. This means that the values and statistical distributions eventually adopted as final set of parameters for the population model are probably not a unique solution, nor are they guaranteed to be the best ones, especially if one adds more observational constraints. They fall however in the range of commonly-accepted values and do provide a decent description of emission properties at the population level, which was our main goal here.

Actually, only modest tuning of the initial model parameters was required to obtain a decent match of predicted and observed flux distributions. This is not surprising because our initial values were informed by previous studies of a subset of the same objects with very similar models. 

Eventually, the most crucial parameters turned out to be: (i) the pulsar birth rate, which linearly sets the normalization of the whole population, but the related parameters are already quite constrained from other observations \citep[exploited in population synthesis efforts such as][]{Lorimer:2006,Faucher-Giguere:2006,Johnston:2020}; (ii) the injection efficiencies in all objects, which determines the maxima in the flux distributions; in the framework of our model, rather high injection efficiencies in \gls{pwne}, at the level $40-100$\% and pretty efficient acceleration in \gls{snrs}, in the $10-30$\% range were required; in both cases, however, this is in line with commonly accepted values, although at the high end for \gls{snrs}; (iii) the initial nebular magnetic field strength in \gls{pwne}, which sets the fraction of the nonthermal particle energy channeled into gamma rays for the younger \gls{pwne} (the rest going into adiabatic losses and synchrotron radiation); to be complete, this should ideally be evaluated together with the time evolution prescription for the nebular field, but we did not engage into this; nevertheless, our adopted value of 50$\,\mu$G is right in the broad range of $\sim10-100\,\mu$G values found in applications of a similar model to various real objects \citep{Zhang:2008,Mayer:2012}; (iv) the circumstellar density distribution in \gls{snrs}: this sets the level of hadronic gamma-ray emission, which is the dominant emission channel for most \gls{snrs} in our model; we ended up with a density distribution with a log-normal mean value of 1\nunit, which is the canonical value for the average \gls{ism}. This mean value is about ten times higher than what was inferred from X-ray \gls{snrs} \citep{Leahy:2017,Leahy:2020}, which is not necessarily surprising since we are comparing two different sets of objects in two different emission bands (X-rays vs. gamma-rays), which is likely to lead to various selection biases (e.g., on age, environment, dynamical state...).

% Synthetic population properties
\section{Synthetic population properties}
\label{popres}

In this section, we present the properties of our synthetic population of \gls{snrs}, \gls{pwne}, and pulsar halos, for the final set of parameters adopted (including several alternative scenarios for halos). The focus is, however, on pulsar-powered objects, \gls{pwne} and halos, because they are the dominant and most numerous sources in the \gls{vhe} sky.

% Population properties
\subsection{Population properties}
\label{popres:props}

Ideally, the synthetic population properties should be described by statistical distributions obtained from a large number of realizations sampling their variance. This would however imply a computation time that is currently out of reach. We present below the results obtained for one realization of the population model that contains a total of 266 \gls{snrs} + 945 \gls{pwne} + 2613 halos = 3824 objects in total. All quantitative prospects, for instance the number of detectable sources, should therefore not be taken as accurate predictions but rather as order of magnitude estimates, subject to statistical fluctuations from the actual sampling of the population for one given set of input parameters (and, maybe more importantly, to changes due to variations in the numerous input parameters).

\begin{figure}[!t]
\begin{center}
\includegraphics[width=0.9\columnwidth]{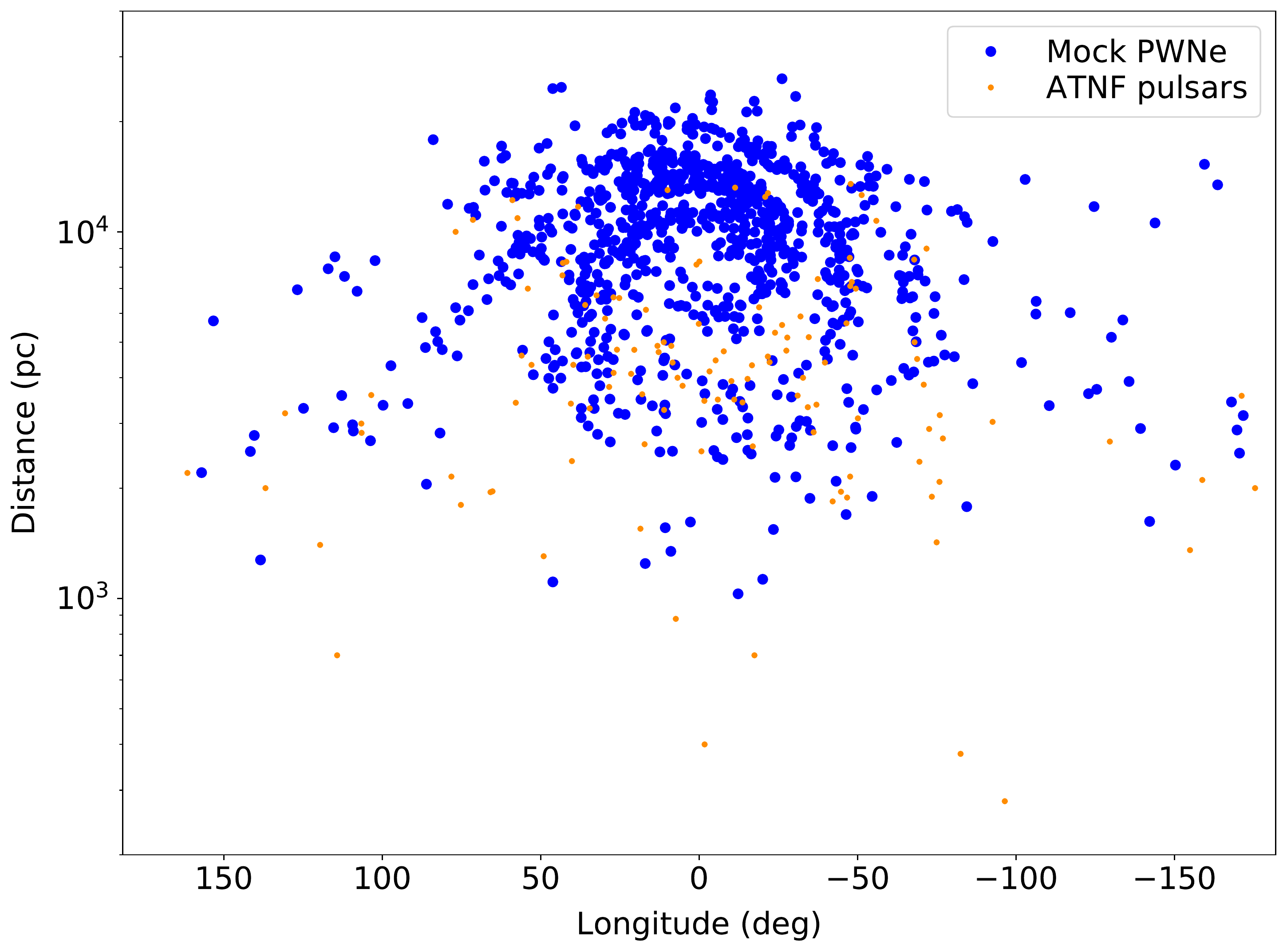}
\includegraphics[width=0.9\columnwidth]{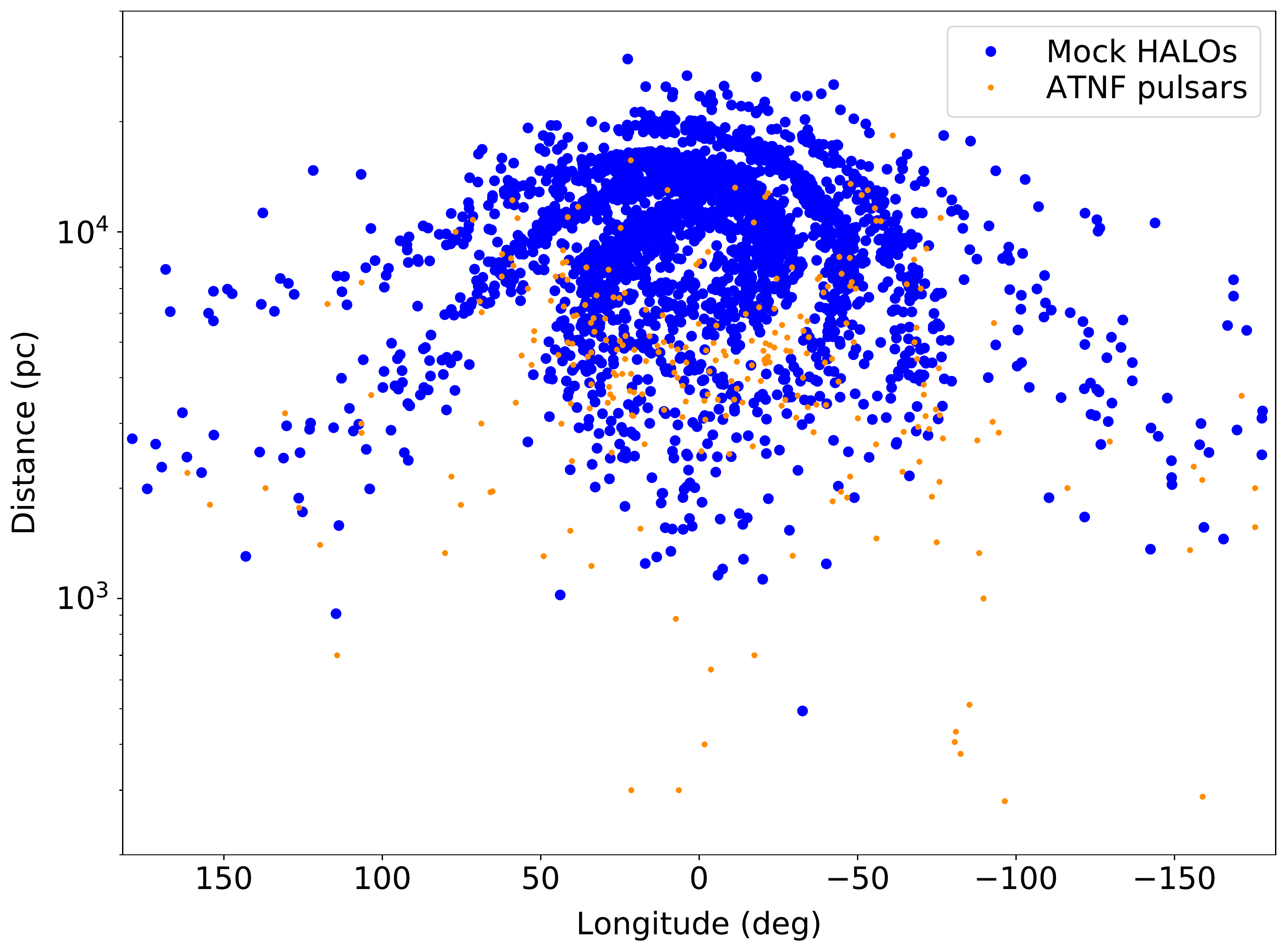}
\caption{Spatial distribution of mock \gls{pwne} and halos in the Galactic plane, for halos with suppressed diffusion region of size 50\,pc and diffusion suppression by a factor 500. Overlaid for comparison and cross-check are the positions of a set of pulsars from the ATNF database within the same age ranges (characteristic ages $<100$\,kyr for \gls{pwne} and $<400$\,kyr for halos).}
\label{fig:res:dists}
\end{center}
\end{figure}

The spatial distribution in the longitude-distance plane are shown in Fig. \ref{fig:res:dists} for \gls{pwne} and halos. We overlaid in the same plane the positions of a set of pulsars selected from the ATNF database after the following filtering: distance $<20$\,kpc, spin-down power $>10^{33}$\punit, and characteristic age below the maximum age allowed for \gls{pwne} or halos in our model (ages $<100$\,kyr for \gls{pwne} and $<400$\,kyr for halos). 

The comparison of the random positions of mock pulsars and actual positions of true pulsars is overall satisfactory, all the more so that the distance estimates for true pulsars are based on a free electron density model that can yield significant uncertainties \citep{Yao:2017}. There is however a deficit of objects within $\lesssim1$\,kpc from the Sun, which is most likely due to the local arm not being included in the spiral arm model from \citet{Faucher-Giguere:2006}. This is particularly apparent in the case of halos, with about four times more known middle-aged pulsars than synthesized. Such a caveat may have important consequences for local observables related to halos, such as the positron flux (see Sect. \ref{popres:posi}).

\begin{figure}[!t]
\begin{center}
\includegraphics[width=0.9\columnwidth]{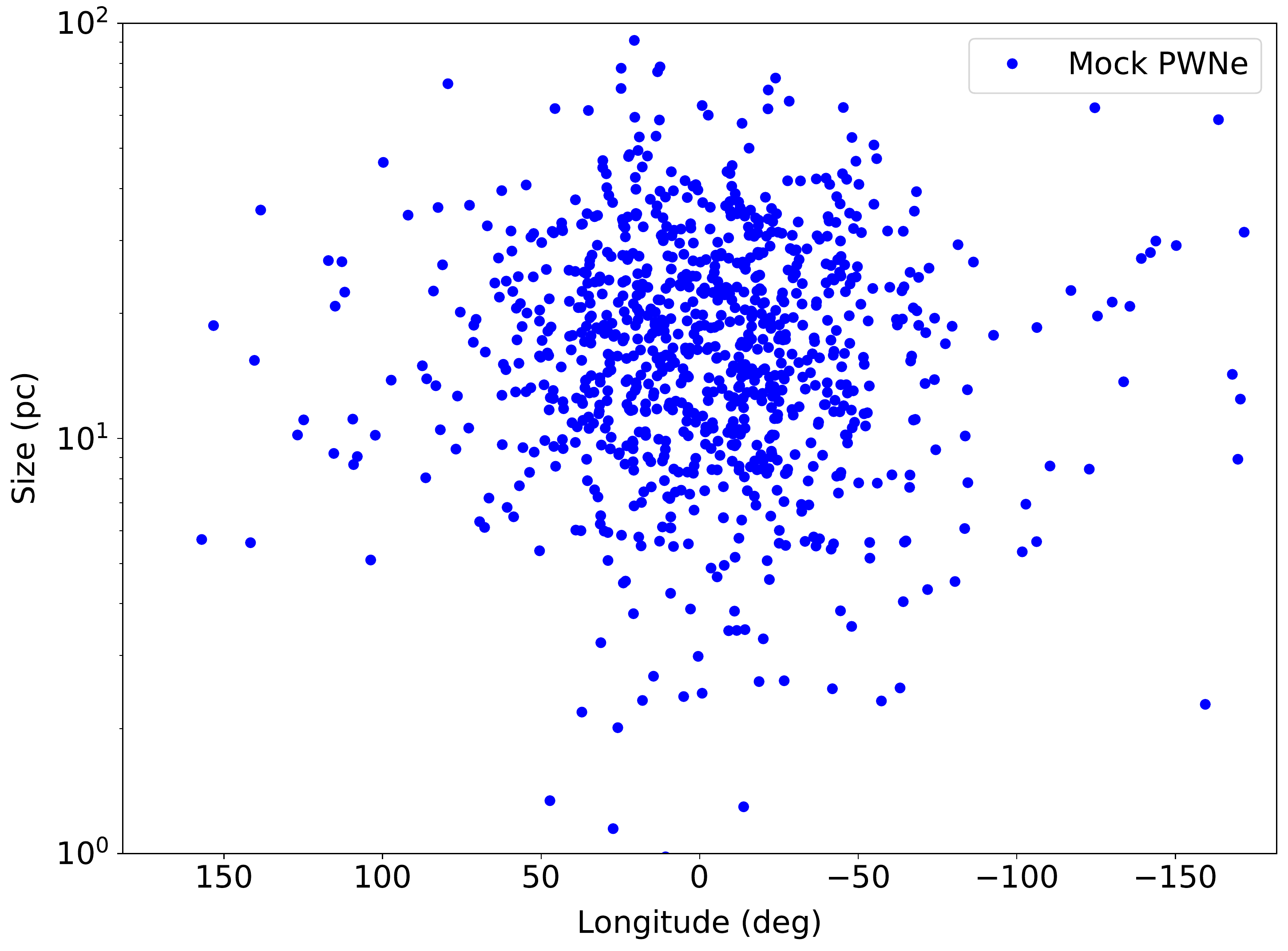}
\includegraphics[width=0.9\columnwidth]{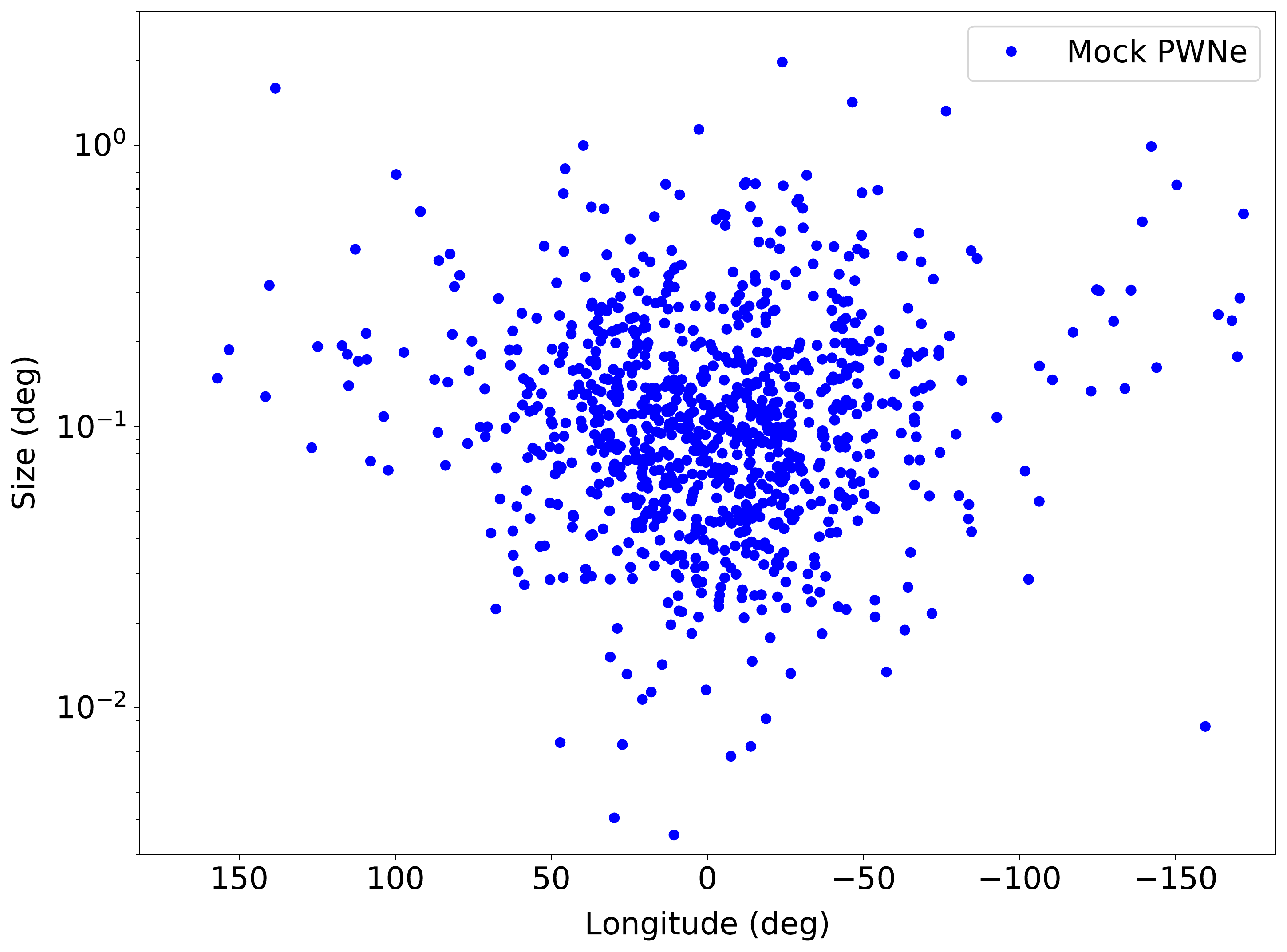}
\caption{Physical and angular sizes for mock \gls{pwne}.}
\label{fig:res:pwnesizes}
\end{center}
\end{figure}

\begin{figure}[!t]
\begin{center}
\includegraphics[width=0.9\columnwidth]{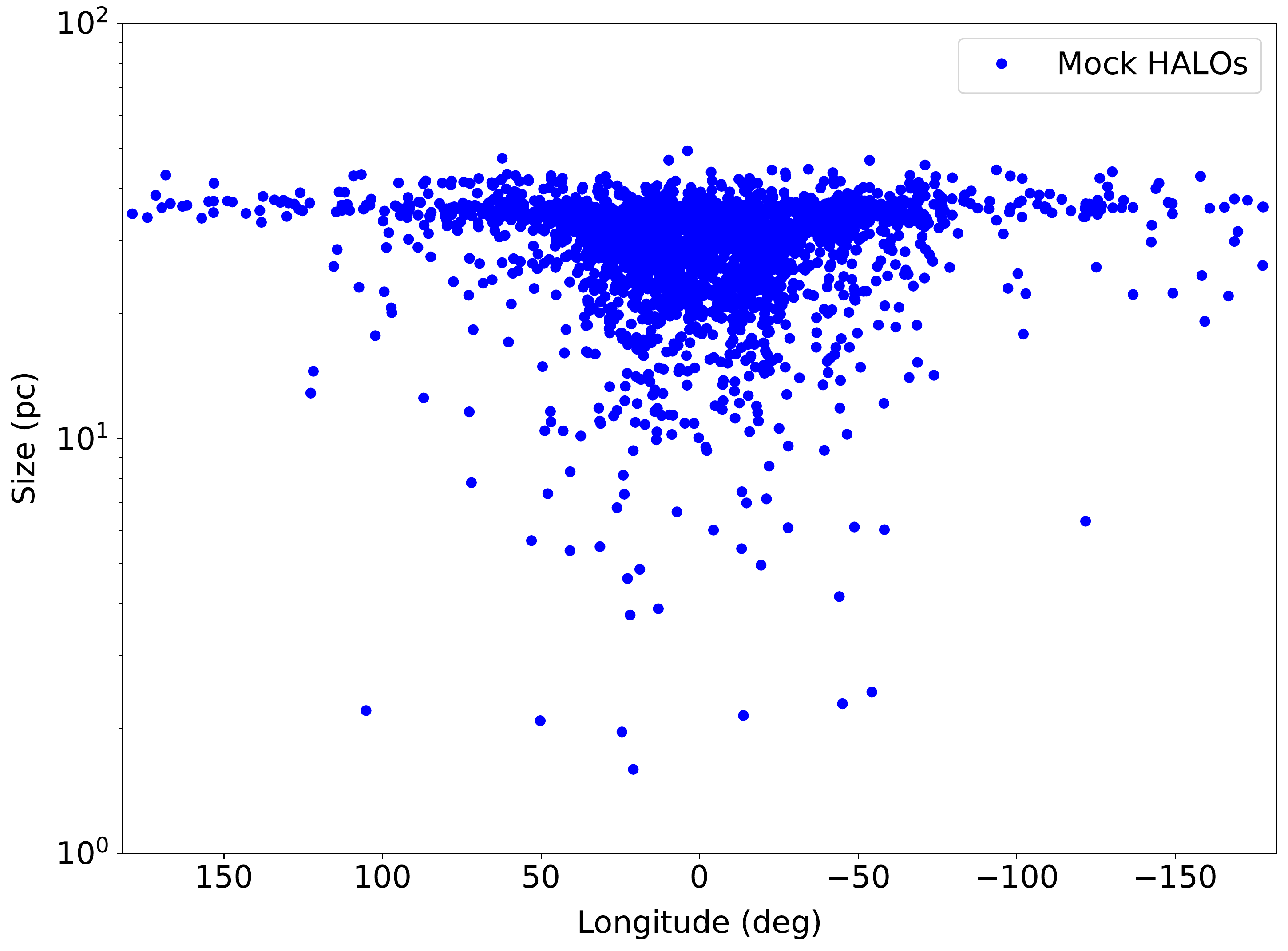}
\includegraphics[width=0.9\columnwidth]{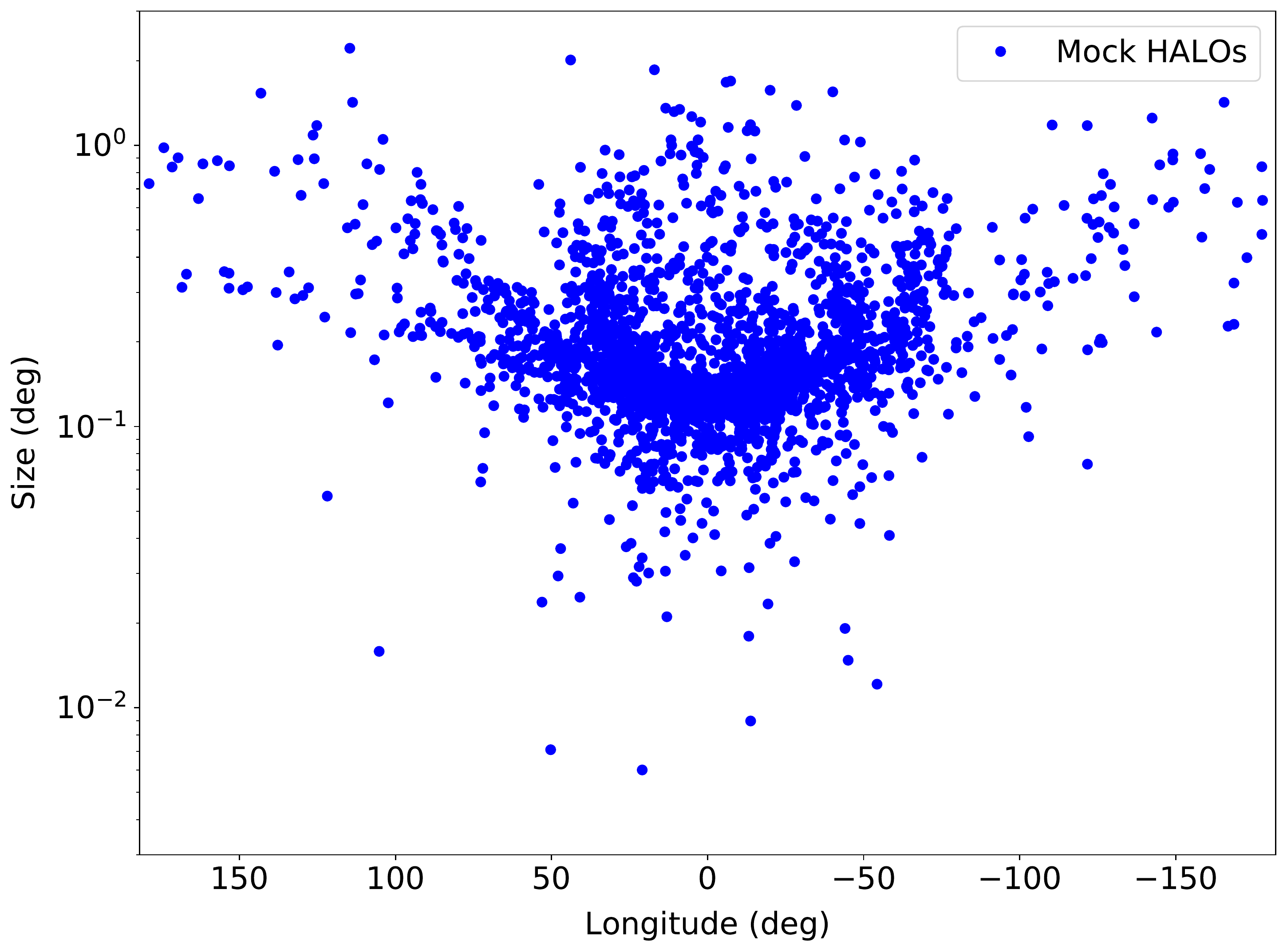}
\caption{Physical and angular sizes for mock halos, modeled with suppressed diffusion region of size 50\,pc and diffusion suppression by a factor 500.}
\label{fig:res:halosizes}
\end{center}
\end{figure}

The physical and angular sizes of \gls{pwne} and halos are displayed in Figs. \ref{fig:res:pwnesizes} and \ref{fig:res:halosizes} (see Sect. \ref{popres:size} for a discussion on the way halo size is computed). The majority of mock \gls{pwne} have physical radii of $\sim5-50$\,pc, which translate into angular extents $\sim0.03-0.3$\deg. The largest objects in the sky reach up to about 100\,pc and the degree scale, similar to the values inferred for HESS J1825-137 \citep{Principe:2020}. 

With our definition for a halo size in the context of \gls{vhe} analyses, the physical sizes of most halos is quite clustered in the $20-50$\,pc range for the halo model setup with 50\,pc suppressed diffusion region. Adopting a more extended confinement region of 80\,pc instead enlarge the size range only modestly, up to 60\,pc. Using a less extended confinement region of 30\,pc gives rise to more subtle effects and a less compact size distribution: halos around weak and young pulsars remain contained in the suppressed diffusion region, with sizes of $20-30$\,pc; older and more powerful pulsar can develop halos extending beyond the suppressed diffusion region, reaching up to 80\,pc. The latter effect is due to significant leakage of particles such that a halo of pairs in excess of the interstellar background can exist over large distances. Eventually, although halos tend to be larger than \gls{pwne} on average, even when allowing for variations of the suppressed diffusion region, there is significant overlap in the angular size distribution of halos and \gls{pwne} in the $\sim0.1-0.3$\deg\ range.

\begin{figure}[!t]
\begin{center}
\includegraphics[width=0.9\columnwidth]{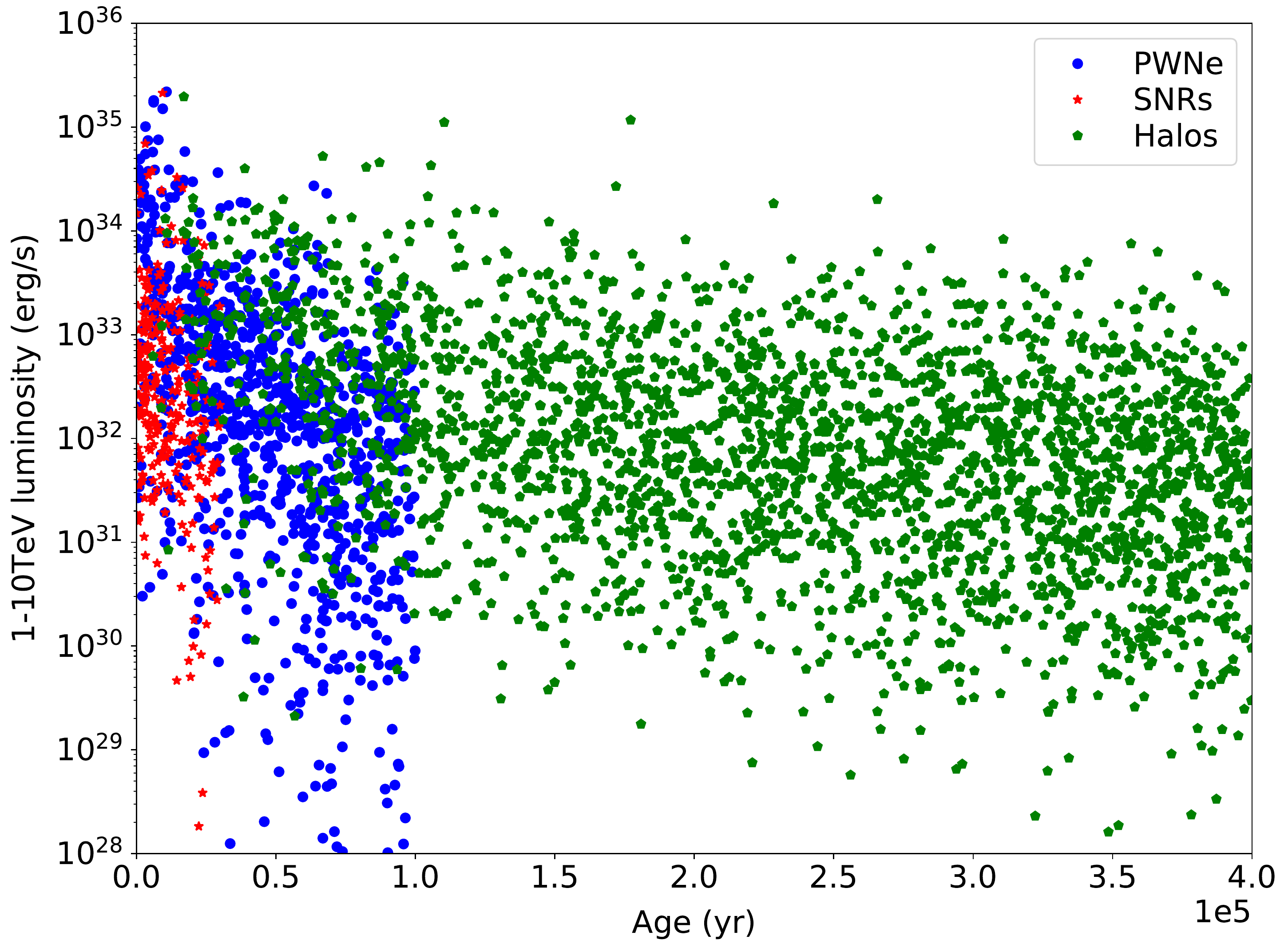}
\includegraphics[width=0.9\columnwidth]{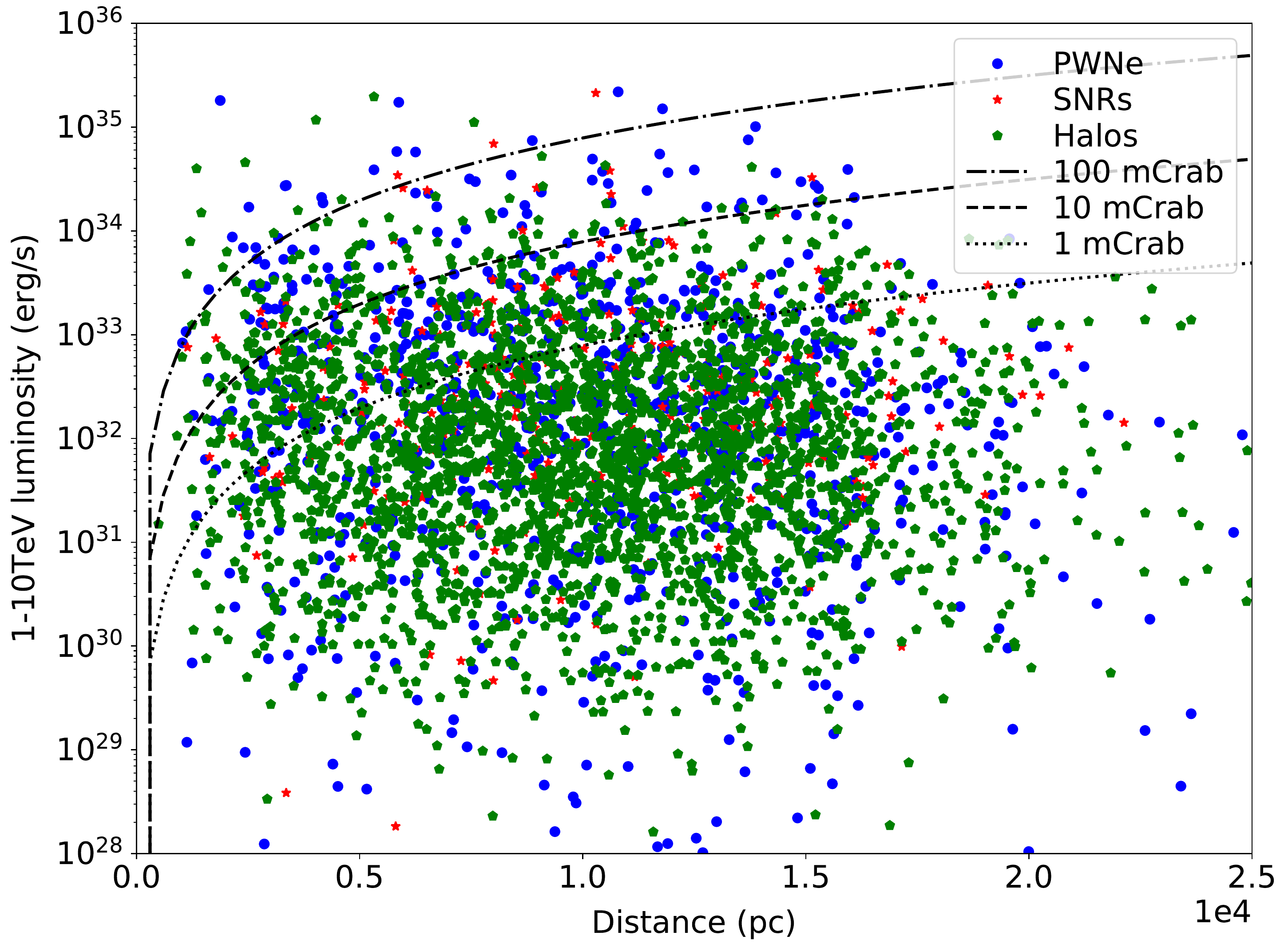}
\includegraphics[width=0.9\columnwidth]{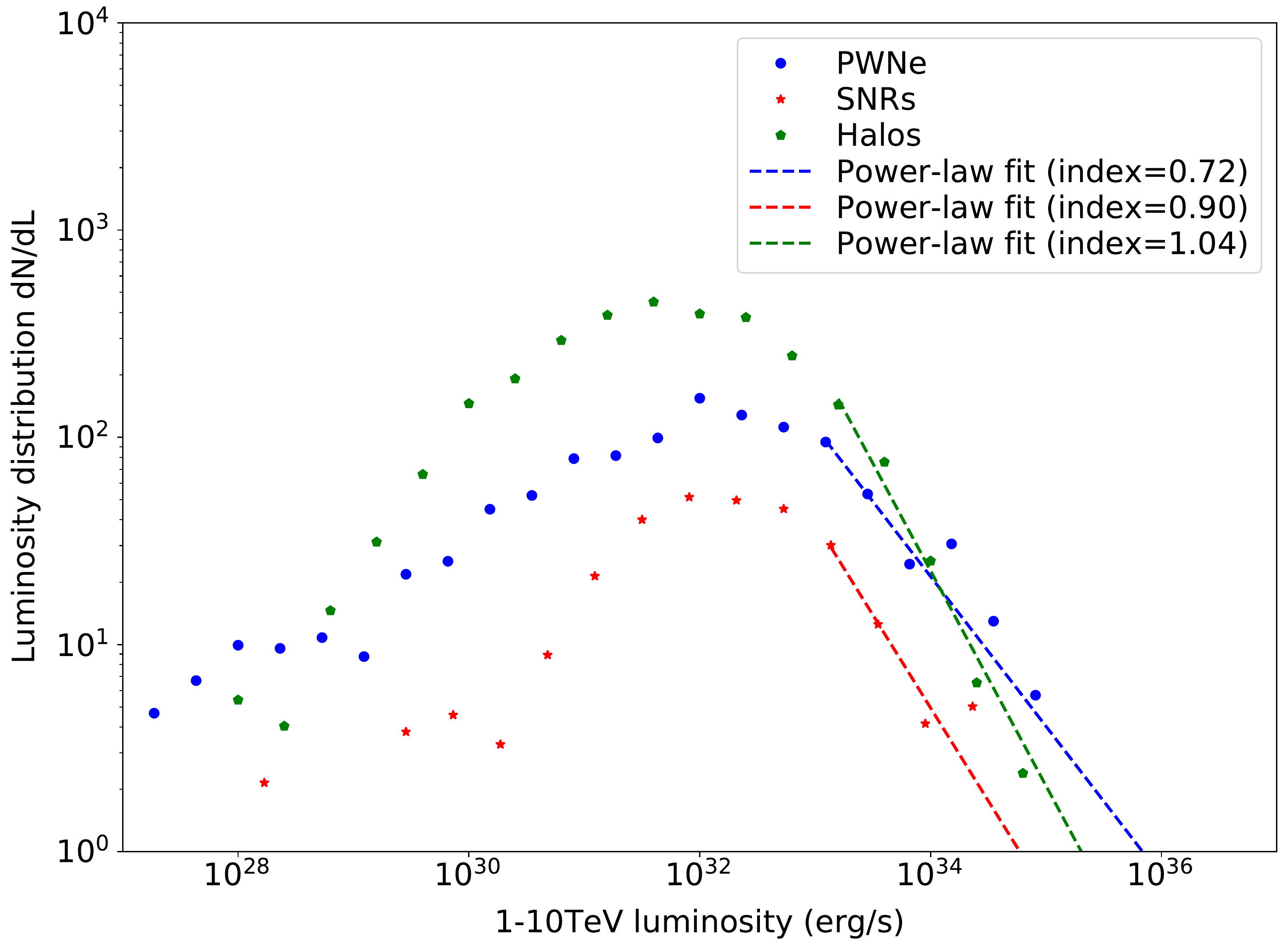}
\caption{Distribution of the $1-10$\tev luminosities for all classes of objects as a function of age, distance, and over the population. Overlaid in the middle panel for comparison are the luminosities corresponding to a 1, 10, or 100\,mCrab flux. In the bottom panel, the dashed lines are power-law fits to the high-luminosity ends of the distributions.}
\label{fig:res:lum}
\end{center}
\end{figure}

Figure \ref{fig:res:lum} displays the $1-10$\tev luminosities of the mock population as a function of age and distance. In the latter case, this is compared to the luminosities corresponding to various reference flux levels to provide a first sense of the accessible fraction of the population\footnote{Throughout the paper, we used as reference spectrum for the Crab nebula the one from \citet{Meyer:2010} as implemented in the gammapy library.}. The top panel shows how halos extend the \gls{vhe} emission of pulsar-powered systems beyond the classical \gls{pwn} stage. Interestingly, with the adopted criterion for the start of the halo phase (see Sect. \ref{model:halo}), there is a significant overlap of the populations of halos and \gls{pwne} in the $10-100$\,kyr range. Such systems are expected to simultaneously harbour both a young relic nebula and a bright halo, most likely offset from each other because of the pulsar motion and reverse-shock crushing of the nebula, and it may be hard to disentangle both components. HESS J1825-137, a very extended source powered by a pulsar with a characteristic age of 21\,kyr, may actually an example of such a system \citep{Principe:2020}. The bottom panel of Fig. \ref{fig:res:lum} shows the luminosity distribution over the populations, together with power-law fits to their high-luminosity ends. Interestingly, these fits reveal distributions that are much flatter than those used or inferred in generic source population synthesis efforts like \citet{Steppa:2020} or \citet{Vecchiotti:2022}. There is a number of reasons for that difference, such as the fact that these works assume that the luminosity and position distributions are independent, but we defer a complete discussion of that point to a subsequent work.

\begin{figure}[!t]
\begin{center}
\includegraphics[width=0.9\columnwidth]{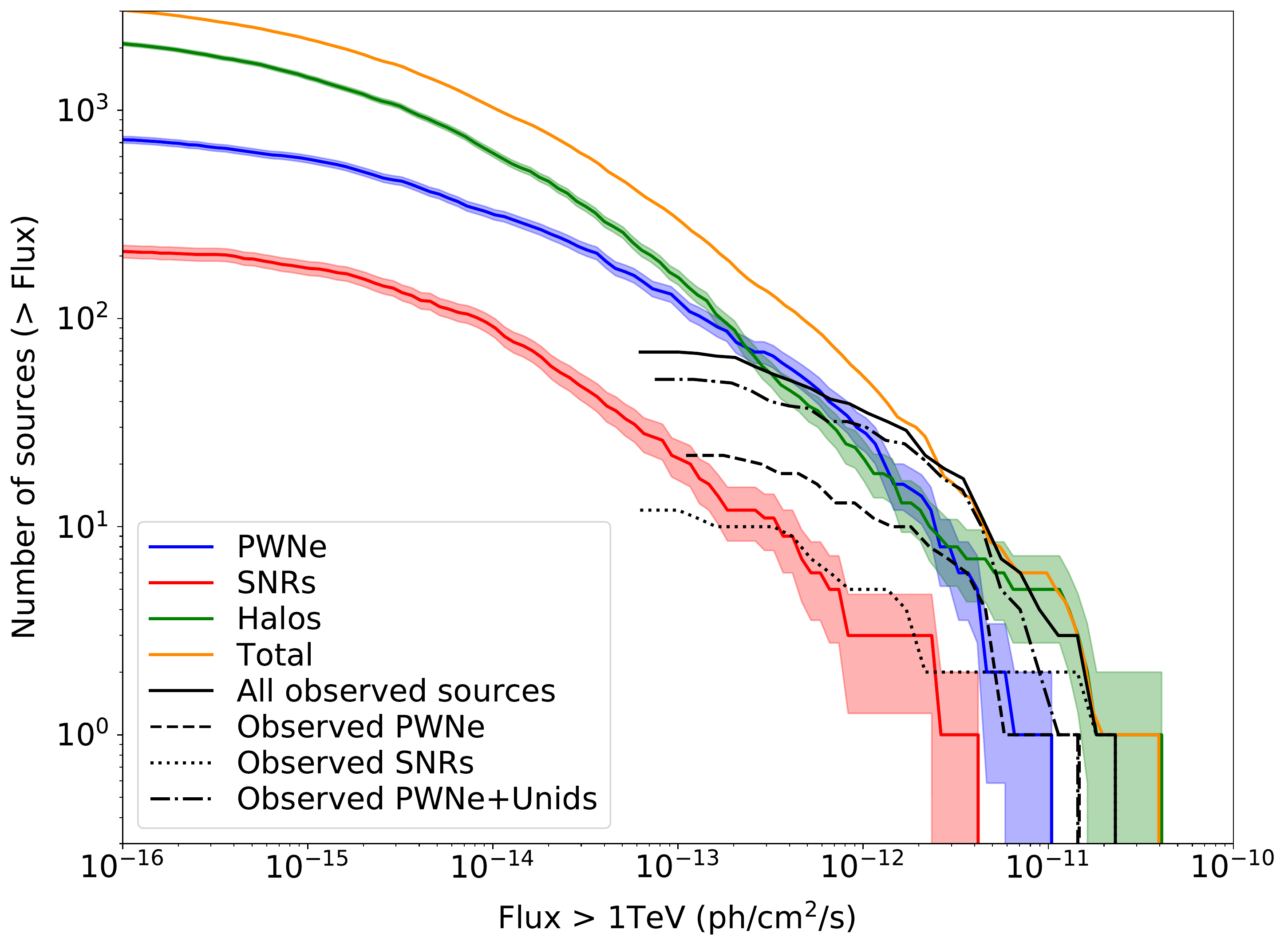}
\includegraphics[width=0.9\columnwidth]{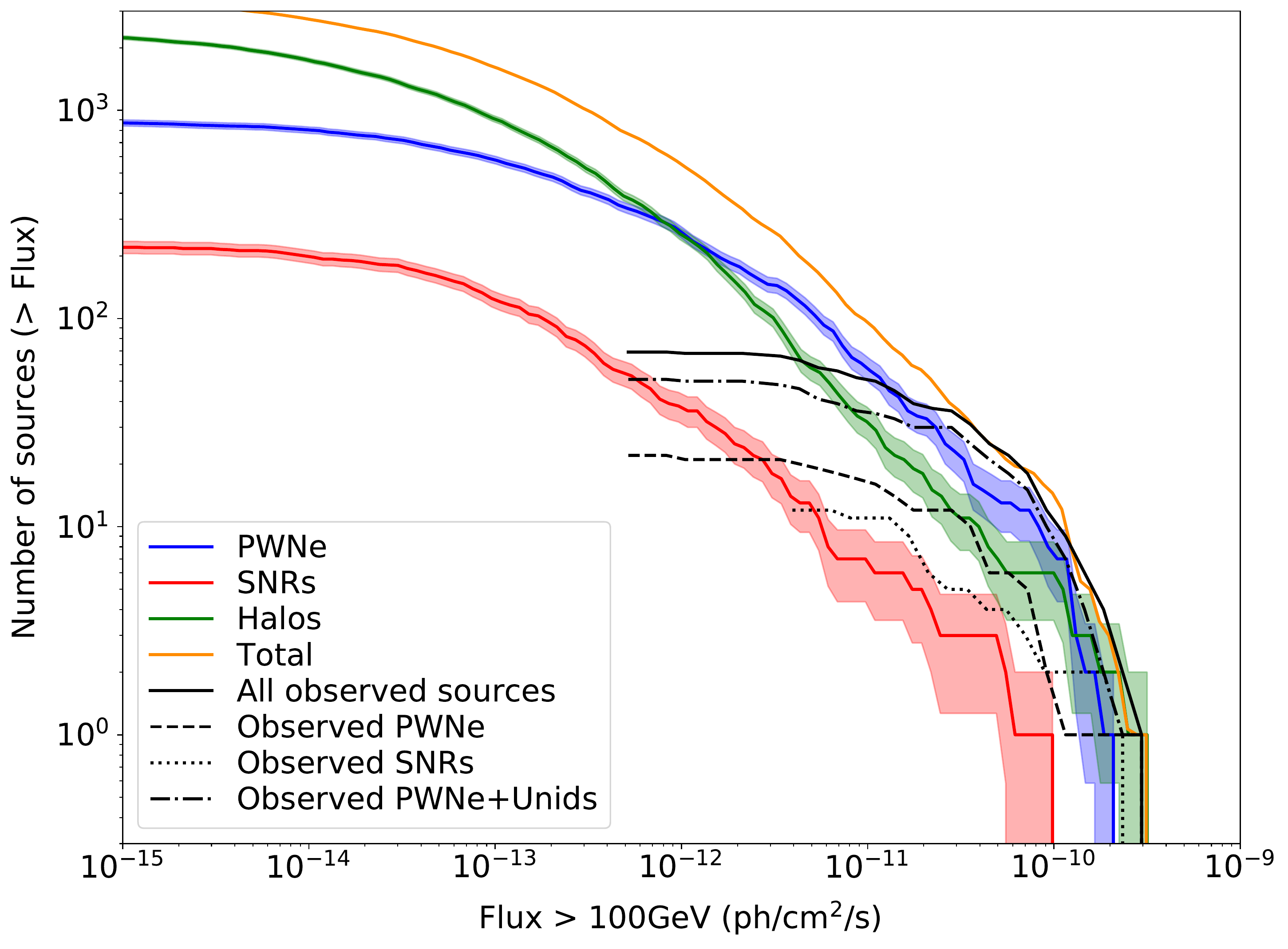}
\caption{Flux distributions above 1\tev and 100\gev of the full population of sources in the footprint of the HGPS, for halos with suppressed diffusion region of size 50\,pc and diffusion suppression by a factor 500. Overlaid in black are the flux distributions of real sources taken from the gamma-cat catalog.}
\label{fig:res:lognlogs500}
\end{center}
\end{figure}

The flux distribution of the full population located in the footprint of the HGPS is presented in the top panel of Fig. \ref{fig:res:lognlogs500}, for a modeling of pulsar halos with a suppressed diffusion region of 50\,pc and diffusion suppression by a factor 500. The population of mock \gls{snrs} reproduces pretty well the distribution of observed objects from the highest fluxes down to about 1\% Crab, except maybe for the presence of two sources shining at the Crab level that this realization of our population model failed to produce. The population of mock \gls{pwne} also matches pretty well the distribution of observed objects from the highest values, at about the Crab level, down to $\sim10\%$ Crab. Below this limit, completeness drops and the observed population becomes an increasingly smaller fraction of the mock population. At $\sim1\%$ Crab, or about the sensitivity limit of current surveys, the number of firmly identified or solid candidate \gls{pwne} is only $\sim20-30$\% of the total number of objects at this flux in the mock population. 

Interestingly, our mock population seems only marginally consistent with \gls{pwne} being the majority of currently unidentified sources above 1\tev, thereby leaving space for another class of objects, possibly halos around a fraction of powerful middle-aged pulsars. The situation is however more contrasted when looking at flux distributions above 100\gev, in the bottom panel of Fig. \ref{fig:res:lognlogs500}, where mock \gls{pwne} are distributed in flux almost like known \gls{pwne}+unidentified sources taken together. This may point to a shortcoming of our model in reproducing the broadband gamma-ray spectra of \gls{pwne}.

With a level of diffusion suppression such as that inferred for the halo around Geminga, the number of halos exceeds that of \gls{pwne} at high fluxes above 10\% Crab. This is mostly a random sampling effect because halos are not expected to be more luminous than the brightest \gls{pwne} on average, as illustrated in Fig. \ref{fig:res:lum}. The flux distributions of halos and \gls{pwne} are similar over the $1-10$\% Crab flux range, which suggests that both classes of objects could be present in similar proportions in the current census of \gls{vhe} sources, if halos are found around a majority of middle-aged pulsars, as assumed by default, and if they can be detected despite their large size. At lower fluxes, halos become more numerous, by about a factor of 2 at 1\,mCrab; we see below, however, that in forthcoming surveys reaching that sensitivity, \gls{pwne} will dominate over halos in number of detectable sources. 

\begin{figure}[!t]
\begin{center}
\includegraphics[width=0.9\columnwidth]{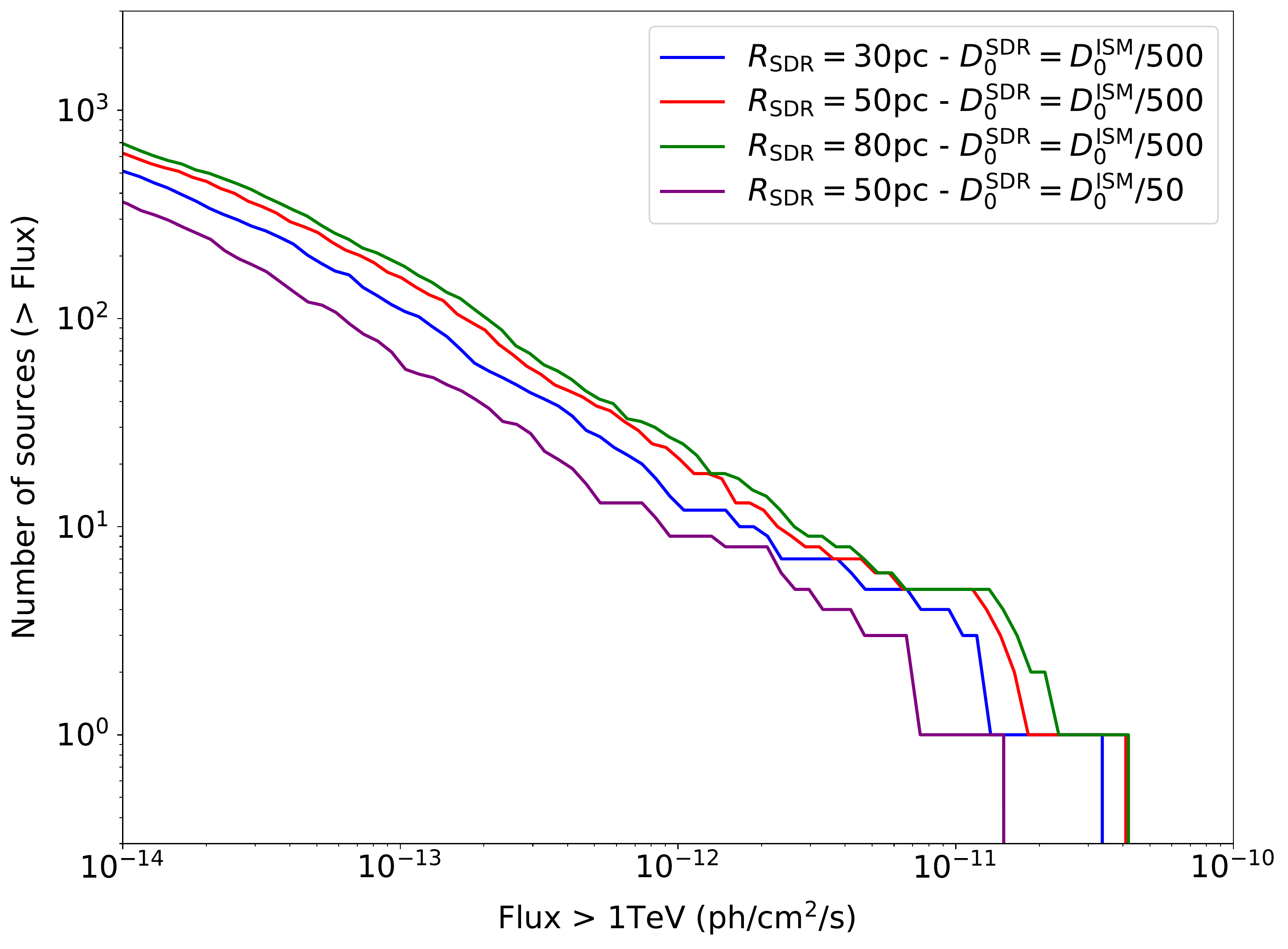}
\caption{Flux distribution above 1\tev of the halo population for different suppressed diffusion region sizes of 30, 50, or 80\,pc but the same diffusion suppression by a factor 500, and for a 50\,pc size and diffusion suppression by a factor 50.}
\label{fig:res:lognlogscomp}
\end{center}
\end{figure}

The flux distribution of the population of halos depends on the extent of the suppressed diffusion region, as illustrated in Fig. \ref{fig:res:lognlogscomp}. The trend is that larger halos tend to be brighter, all other things being equal. At the sensitivity limit of current surveys, about 10\,mCrab, or at the level to be reached in future surveys, about 1\,mCrab, the number of halos with large extents $\gtrsim$80\,pc is $50-100$\% larger than the number of halos with small extents $\lesssim$30\,pc. This difference should in principle be discernible when the number of detectable sources becomes sufficient, as we discuss more quantitatively in the next section. The flux distribution of halos depends more strongly on the level of diffusion suppression. Modeling halos with shallower diffusion suppression, consistent with the level inferred for PSR B0656+14 instead of that for PSR J0633+1746, results in $2-3$ times fewer objects at any given flux. As discussed below, the prospects for detection are even more degraded because such halos also have a larger angular extent. 

Overall, our population model does provide a satisfactory description of the currently observed population at fluxes above $5-10$\% Crab. In the next section, we demonstrate that the model can also account fairly well for the properties of observed sources at lower fluxes, by assessing the detectable fraction of the population in existing surveys and comparing it to the actual results. The predicted \gls{pwne} population does not saturate the flux distribution of known objects with $>1$\tev fluxes above $5-10$\% Crab, which leaves room for another class of emitters as likely counterparts to the currently unidentified sources. Pulsar halos seem to be a likely possibility.

% Detectable fraction of the population
\subsection{Detectable fraction of the population}
\label{popres:det}

\begin{table*}[ht!]
\centering
\begin{tabular}{| c | c c c c c | c |}
\hline
\celltspace Survey & \gls{snrs} & \gls{pwne} & Halos (30\,pc) & Halos (50\,pc) & Halos (80\,pc) & Total \\
\hline
\celltspace \textit{Total in population} & 266 & 945 & 2613 & 2613 & 2613 & 3824 \\
\hline
\celltspace H.E.S.S. (HGPS) & 8 & 54 & 17 & 23 & 23 & 79, 85, or 85 \\
\hline
\celltspace HAWC (3HWC) & 2 & 19  & 10 & 15 & 16 & 31, 36, or 37 \\
\hline
\celltspace CTA (GPS) & 30 & 171 & 43 & 74 & 103 & 244, 275, or 304 \\
\celltspace CTA (GPS+) & 44 & 261 & 103 & 166 & 217 & 408, 471, or 522 \\
\hline
\end{tabular}
\caption{Number of detectable sources in past or forthcoming surveys, obtained for one single realization of the population model. The prospects for halos are given for three alternative modeling, with different radii for the suppressed diffusion regions of 30, 50, or 80\,pc, and diffusion suppression by  factor 500 (see Table \ref{tab:modpars}). Prospects for \gls{cta} are also given for a variant of the survey, GPS+, with sensitivity improvement by a factor of 2 with respect to the values published in \citet{Acharya:2019}.}
\label{tab:detsrcs} 
\end{table*}

We assessed the fraction of the mock population that should have been detected in existing or past surveys or ought to be detected in forthcoming surveys. For that purpose, we used very simple criteria for the detectability, typically that flux is above a certain threshold characteristic of the survey in question. Actual data analysis is otherwise more complex, as thoroughly exposed for instance in the case of \gls{hgps} \citep{Abdalla:2018a}.

We considered the \gls{hgps} \citep{Abdalla:2018a}, the HAWC 1523-day survey from which the 3HWC catalog was derived \citep{Albert:2020}, and the future CTA Galactic Plane Survey, hereafter GPS \citep{Acharya:2019}, for which we adopted the following criteria for detectability:
\begin{enumerate}
\item H.E.S.S.: Integrated $>1$\tev photon flux above 10\,mCrab, for a source located within $[-100\deg,70\deg]$ in longitude and $[-2\deg,2\deg]$ in latitude; sensitivity degradation as a result of source extension was implemented following Eq. 28 in \citet{Abdalla:2018a}, assuming a 0.08\deg\ point-spread function, with a limit of 0.7\deg\ in radius beyond which any source is considered undetectable.
\item HAWC: For a source located within $[-20\deg,60\deg]$ in declination, 7\tev photon flux density above $3 \times 10^{-15}$\,ph\,cm$^{-2}$\,s$^{-1}$\,TeV$^{-1}$ at best, degrading up to $7 \times 10^{-15}$\,ph\,cm$^{-2}$\,s$^{-1}$\,TeV$^{-1}$ following a parabolic dependence on declination with minimum at 19\deg; sensitivity degradation as a result of source extension was implemented as above assuming a 0.2\deg\ point-spread function.
\item CTA: Integrated $>125$\gev photon flux above 1.8, 2.7, 3.8, 3.1, or 2.6\,mCrab, for a source located in the longitude range $[-60\deg,60\deg]$, $[60\deg,150\deg]$, $[150\deg,-150\deg]$, $[-150\deg,-120\deg]$, and $[-120\deg,-60\deg]$, defined respectively as Inner, Cygnus+Perseus, Anticenter, and Vela+Carina in \citet{Acharya:2019}; the latitude range for all segments was taken to be [-3\deg,3\deg], and sensitivity degradation as a result of source extension was computed as above assuming a 0.07\deg\ point-spread function, with a (somewhat arbitrary) limit of 2.0\deg\ in radius.
\end{enumerate}
The threshold used for the HGPS is an average over the $\sim0.5-1.5$\% Crab values reached over most of the surveyed area, which however corresponds to optimistic sensitivities to isolated point sources that are recognized overestimates of the actual sensitivities in \citet{Abdalla:2018a}. Conversely, the $\sim2-3$\,mCrab sensitivities presented for the CTA GPS correspond to estimates that were most likely refined since the publication of \citet{Acharya:2019}, as the final design of the instrument was settled and its performances were investigated more in depth \citep{Remy:2022}. The Southern Wide-field Gamma-ray Observatory will no doubt advance the census and knowledge of extended TeV sources, including pulsar halos, but the design of the project is not yet sufficiently settled for quantitative prospects to be derived \citep{Hinton:2022}.

The angular extent of a source enters in the above criteria for detection, as it implies a mixing of the source signal with more background (instrumental or astrophysical in origin, although here we consider only the effect of instrumental background). Source extent is easily defined for \gls{snrs} and \gls{pwne}, as the forward shock and outer nebula radius, respectively, but is more complicated for halos. We describe in Sect. \ref{popres:size} how halo size is computed and how alternative definitions impact the number of detectable halos.

From one realization of the population model, and based on the above detectability criteria, the number of detectable sources we obtain are reported in Table \ref{tab:detsrcs}. Out of a total of $\sim3800$ objects, it is predicted that the H.E.S.S. survey of the Galactic plane should have detected about 80, with little dependence on how pulsar halos are modeled in terms of extent. The majority of sources accessible to the survey would be \gls{pwne} (about 50), followed by halos (about 20), and then \gls{snrs} (about ten). This is consistent with the actual outcome of the survey \citet{Abdalla:2018a} if pulsar-powered objects do constitute the majority of currently unidentified or unassociated sources (47 out of 78).

The predicted number of detectable sources in the HAWC observations used in the 3HWC catalog making is about 30-40, split almost evenly between \gls{pwne} and halos and with a low number of \gls{snrs}. This undershoots the actual number of detected sources, which is 65. The latter number however comprises 17 objects that are not well separated from neighboring sources and may just be secondary local maxima resulting from statistical fluctuations \citep{Albert:2020}. The number of truly distinct sources in the 3HWC catalog could therefore be as low as 48, which becomes more consistent with our model prediction within Poisson fluctuations. 

Last, prospects for detection with the upcoming \gls{cta} are very promising, with as many as 30 \gls{snrs}, 170 \gls{pwne}, and $40-100$ pulsar halos accessible to the GPS (with an additional $\sim10-20$\% uncertainty on the latter figure owing to the choice of a typical extent for detectability assessment; see Sect. \ref{popres:size}). If the survey sensitivities along the plane are better than those published in \citet{Acharya:2019}, by a factor of $\sim2$ leading to $\sim1$\,mCrab sensitivity in the innermost regions, the number of detectable sources increases significantly, by about 50\% for \gls{snrs} and \gls{pwne}, and by more than 100\% for halos (the stronger increase for halos reflecting the fact that a larger fraction of the population reside at low fluxes). 

It is interesting to compare these numbers with the prospects presented in \citet{Remy:2022} from a simulation and analysis of CTA survey observations in conditions close to our assumed GPS+ scenario and with a different model for source populations (including more source classes like binaries and interacting \gls{snrs}, excluding pulsar halos, and using alternative prescriptions for \gls{pwne}). The numbers of detectable simulated sources is on the order of 290 \gls{pwne}, 40 young shell \gls{snrs}, and 100 already known objects that include mostly \gls{pwne} and unidentified sources. We obtain 260 \gls{pwne}, 40 \gls{snrs}, and $100-200$ pulsar halos. This is rather consistent because the source model in \citet{Remy:2022} was tuned to have the mock population of \gls{pwne} account for the large fraction of currently unidentified sources, whereas we did not enforce such a requirement.

In this assessment, the recent H.E.S.S. and HAWC surveys are little sensitive to the extent of the suppressed diffusion region in halos. The small number of such objects accessible to these current instruments is too low to discern the effect that larger halos are on average brighter. In contrast, the better prospects for detection offered by \gls{cta}, with about $50-100$ halos or more, should in principle make it possible to study such trends at a population scale. This illustrates the potential of \gls{cta} for a deeper investigation of pulsar halos, although it is unclear which fraction of these detectable halos could actually be identified as such, and which fraction could lead to solid insight into their physics. 

Using a more moderate diffusion suppression in halos, with a factor 50 instead of 500 (that is more representative of what is inferred for the halo of PSR B0656+14 than for that of PSR J0633+1746), prospects for detection of this class of objects collapse. They fall down to a handful for the H.E.S.S. and HAWC surveys, and less than 20 for \gls{cta}. The main reason is that the higher diffusion coefficient in the close vicinity of the pulsars results in much more extended emission structures, which thus evade detection. On the other hand, we are probably reaching here the limit of our simple method to assess detectability and a more appropriate treatment would be needed, taking into account the full spectro-morphological properties on the halo and the actual energy-dependent sensitivity of the instrument over a large band.

\begin{figure}[!t]
\begin{center}
\includegraphics[width=0.9\columnwidth]{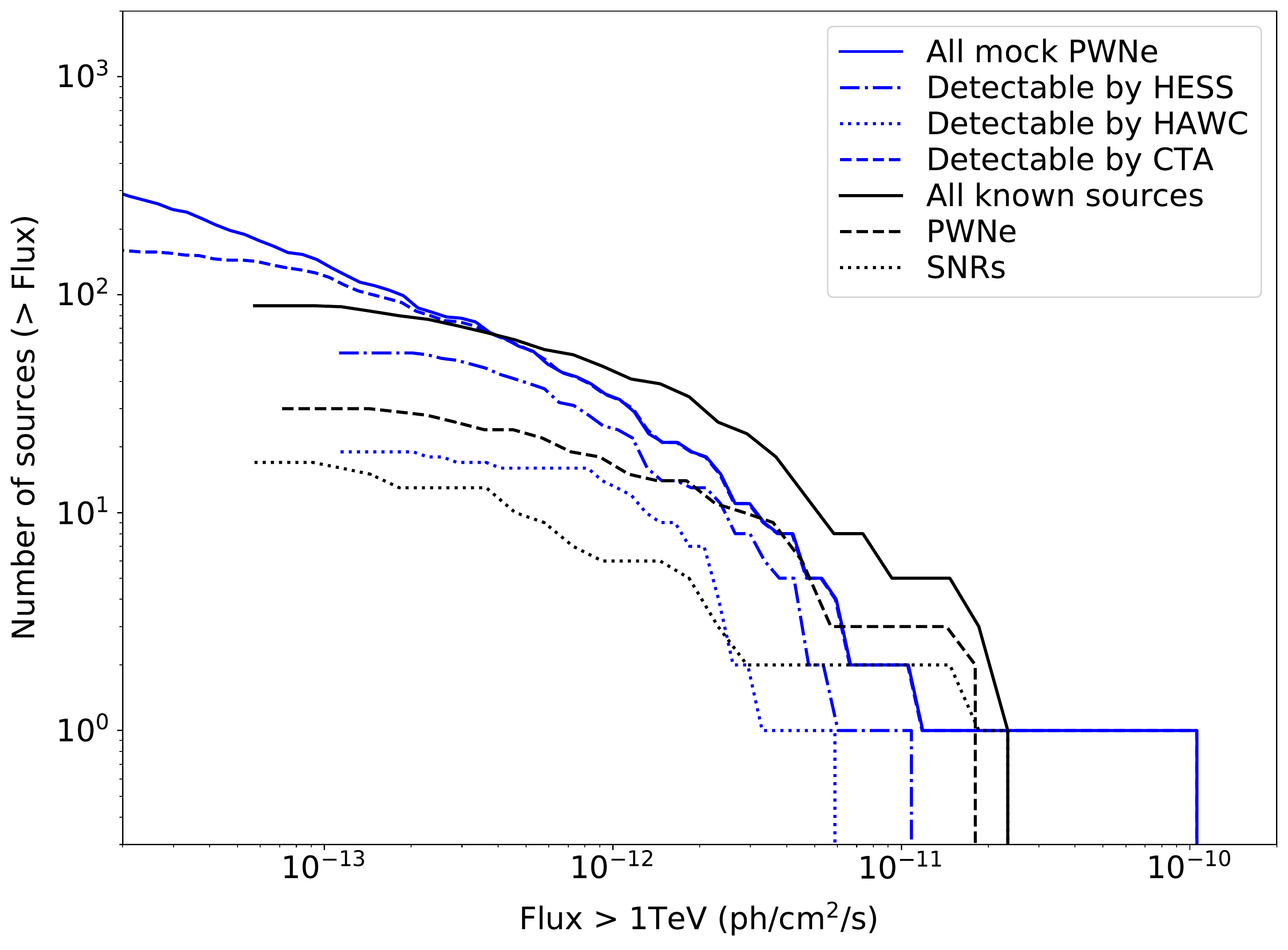}
\includegraphics[width=0.9\columnwidth]{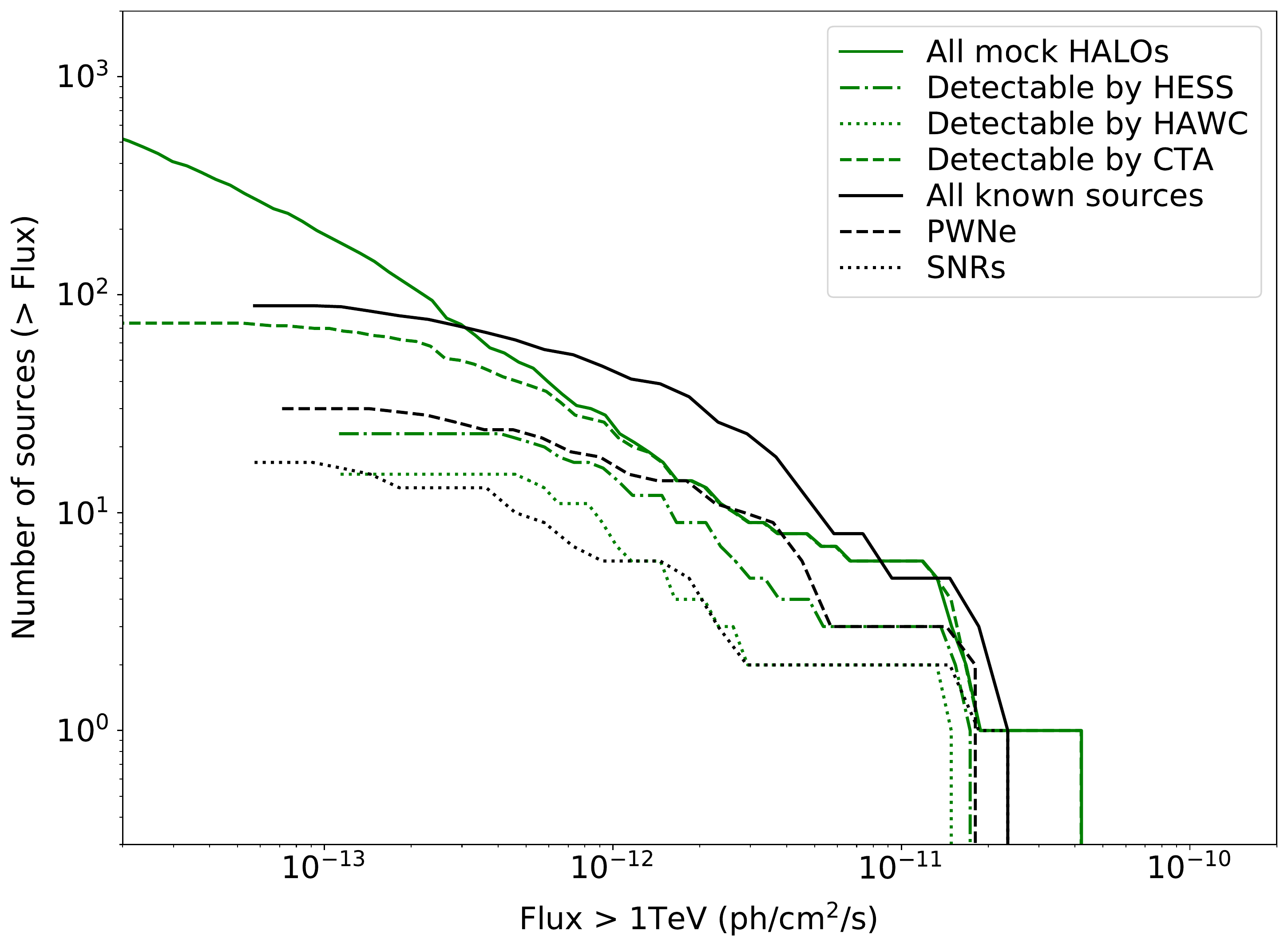}
\caption{Flux distribution of the detectable fractions of the populations of \gls{pwne} and pulsar halos, modeled with suppressed diffusion region of size 50\,pc and diffusion suppression by a factor 500. For comparison, the distributions of all mock and known sources over the full sky are displayed.}
\label{fig:res:detsrcs}
\end{center}
\end{figure}

Figure \ref{fig:res:detsrcs} displays the flux distributions of detectable \gls{pwne} and halos in the existing or future surveys, compared to the distributions of all mock and known sources over the full sky. The plot nicely illustrates the step that will be made possible with \gls{cta} in probing with a high degree of completeness the populations of pulsar-powered sources over about two orders of magnitude in flux. Such prospects suggest or rather confirm that forthcoming surveys will have to deal with a high level of source confusion, especially when the complexities of realistic morphologies are folded in. The majority of the sources included in our population are found in the innermost $100-120$\deg\ of the Galactic plane, which means for \gls{cta} an average $2-4$ detectable sources per degree in longitude, with extensions typically of $\sim0.1-0.3\deg$ but reaching up to the degree scale.

% Impact of size estimates for pulsar halos
\subsection{Impact of size estimates for pulsar halos}
\label{popres:size}

While the physical extent of \gls{snrs} and \gls{pwne} can be unambiguously defined in the framework of our models, as the forward shock and outer nebula radius, respectively, this is more problematic for halos since they do not have a well-defined external boundary and feature a strongly energy-dependent morphology (the latter being also true for \gls{pwne}, but our simple model does not allow us to properly describe this effect). Since halos can be quite extended emission structures, reaching beyond the angular resolution of \gls{vhe} instruments, the way their size is defined can have strong consequences when assessing their detectability.

The only characteristic scale of a halo, the extent of the suppressed diffusion region, is not quite relevant. Depending on the parameters of the system (such as suppressed diffusion coefficient and injection efficiency, pulsar power and age), the resulting pair halo will constitute an overdensity on top of interstellar background up to a distance that varies with energy, and in some cases may not even yield any significant excess at all. Therefore, we defined the typical extent of a halo at any given energy as the radius where the pair density equals that of the interstellar background. This does not come without ambiguity, however, because the density of cosmic-ray leptons at very high energies $\gtrsim100$\gev can be expected to fluctuate significantly, in space and time, as illustrated in \citet{Porter:2019}. In the absence of information about its value at any position in the Galaxy, we used as typical interstellar lepton background that inferred in our local neighborhood, which we approximated in the $\sim0.1-10$\tev range as a power law with slope 3.18 and flux normalization $E^3 \times \Phi_e(E) = 100$\,GeV\,m$^{-2}$\,s$^{-1}$\,sr$^{-1}$ at 1\tev \citep{Aguilar:2019b}.

A second problem then is the choice of the typical extent to use when assessing detectability by a given instrument with the simple approach introduced in Sect. \ref{popres:det}. Broadband detectability assessment should in principle exploit the complete spectro-morphological properties of each source, given the actual spectral sensitivity of the survey at its position in the plane. Such a dedicated approach is however intractable in the context of our population synthesis, with a few thousands objects. We need instead one effective extent that would yield a good estimate of the detectability over the few hundreds of GeV to few tens of TeV band. From a study of the detectability of halos with CTA that some of us are involved in \citep{Eckner:2022}, it seems that the optimum energy for the detection of halos is in the few TeV range, on average: at lower energies, the halo is brighter but it is also more extended and the instrumental sensitivity degrades; at higher energies, the halo is more compact but it is also dimmer and sensitivity starts degrading too. 

Ultimately, we used two definitions for a halo angular size: the 68\% or 95\% containment radius of the 3\tev flux, within the region where the halo pair density is above the local interstellar lepton background at 10\tev. Variations of the containment fraction or reference energies result in changes by 20\% at most, as illustrated in Table \ref{tab:halosizes}. The 68\% containment fraction is applicable to detectability assessment, because it implies a higher surface brightness over a smaller patch of sky, which is more favorable to detection. The 95\% containment fraction is applicable to a general purpose and permits a more consistent comparison with \gls{snrs} and \gls{pwne} (whose sizes encapsulate 100\% of the emission).

\begin{table}[t]
\centering
\begin{tabular}{| c | c | c | c |}
\hline
\celltspace $f_c$ & $E_{e}^{over}$ (TeV) & $E_{\gamma}^{ref}$ (TeV) & $N_{\textrm{HALOs}}$\\
\hline
\celltspace 0.68 & 1 & 1 & 70 \\
\celltspace 0.95 & 1 & 1 & 56 \\
\celltspace 0.68 & 10 & 1 & 69 \\
\celltspace 0.95 & 10 & 1 & 58 \\
\celltspace 0.68 & 10 & 3 & 74 \\
\celltspace 0.95 & 10 & 3 & 63 \\
\celltspace 0.68 & 10 & 10 & 82 \\
\celltspace 0.95 & 10 & 10 & 70 \\
\celltspace 0.68 & 100 & 10 & 83 \\
\celltspace 0.95 & 100 & 10 & 71 \\
\hline
\end{tabular}
\caption{Number of detectable halos in the CTA survey, $N_{\textrm{HALOs}}$, depending on how their typical size is defined (lepton energy $E_{e}^{over}$ at which the extent of pair halo overdensity on top of interstellar background is evaluated, reference gamma-ray energy $E_{\gamma}^{ref}$ at which the flux profile is taken, and containment fraction $f_c$ of the latter). The numbers correspond to a population of halos with suppressed diffusion region of size 50\,pc and diffusion suppression by a factor 500.}
\label{tab:halosizes} 
\end{table}

% Diffuse emission from the unresolved population
\subsection{Diffuse emission from the unresolved population}
\label{popres:diff}

\begin{figure}[!t]
\begin{center}
\includegraphics[width=0.9\columnwidth]{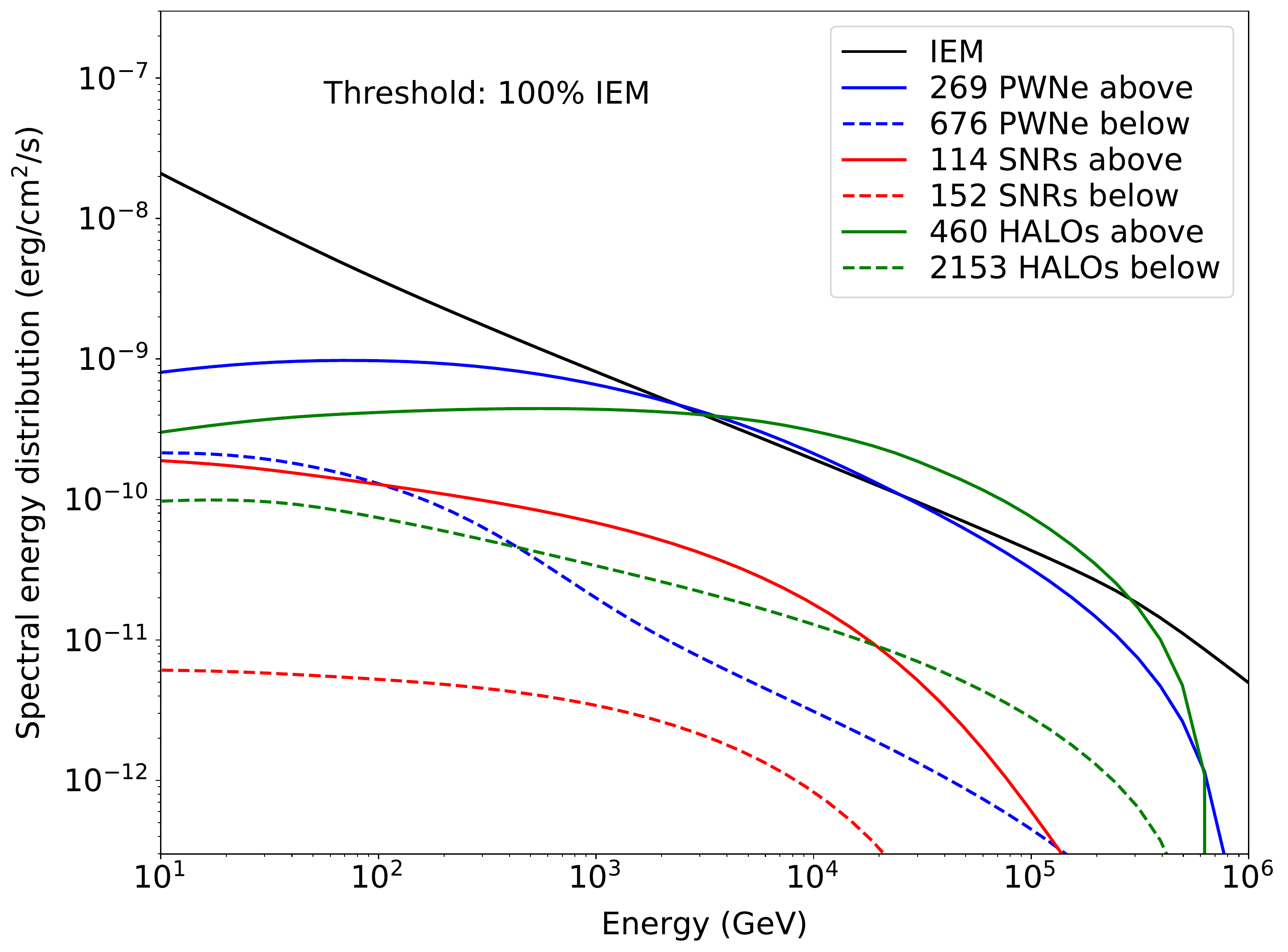}
\caption{Cumulative emission from objects that are individually brighter or fainter than coincident interstellar radiation, in the $1-10$\tev range, compared with the ``Base Max'' interstellar emission model from \citet{DeLaTorreLuque:2022} integrated over the whole sky (in black, labeled ``IEM'').}
\label{fig:res:diffism}
\end{center}
\end{figure}

Objects that are not detectable individually in existing or future surveys can be expected to give rise to some diffuse emission. In this section, we assess the spectral properties of the total emission from the unresolved population of each class of objects, and compare it to a model for the interstellar emission powered by the large-scale population of Galactic \gls{crs} interacting with the \gls{ism}. For the latter, we use the so-called ``Base Max'' model from \citet{DeLaTorreLuque:2022}, which is representative of the conventional interstellar transport scenario. Among the different model setups explored in that paper, the ``Base Max'' variant yields the smallest intensities in the innermost regions of the Galaxy and over the $0.1-10$\tev range. This minimal prediction already seems disfavored at TeV energies by {\it Fermi}-LAT and ARGO-YBG measurements, but diffuse emission from unresolved sources may have a non-negligible in that range, as we discuss below.

We first compare each population of sources to the interstellar emission, independently of the detectability by any instrument. For each object, we integrate the interstellar emission within its angular extent and over the $1-10$\tev band, and compare it to the predicted flux from the object itself, within the same angular region. We then sum up the contributions for all objects with emissions above or below the interstellar radiation. The result is presented in Fig. \ref{fig:res:diffism}. Only one third of \gls{pwne} and one fifth of halos are brighter than coincident interstellar emission in the TeV range. This emission integrated over the full sky is one to two orders of magnitude above that of fainter objects, for any class of sources. In the case of \gls{pwne} and halos, it is comparable to or exceeds interstellar radiation over $2-200$\tev. This can have interesting consequences on observations of external galaxies, such as M31, where the population of pulsar-powered sources may rival in intensity with interstellar emission and bias the interpretation of diffuse emission in terms of \gls{cr} transport.

\begin{figure}[!t]
\begin{center}
\includegraphics[width=0.9\columnwidth]{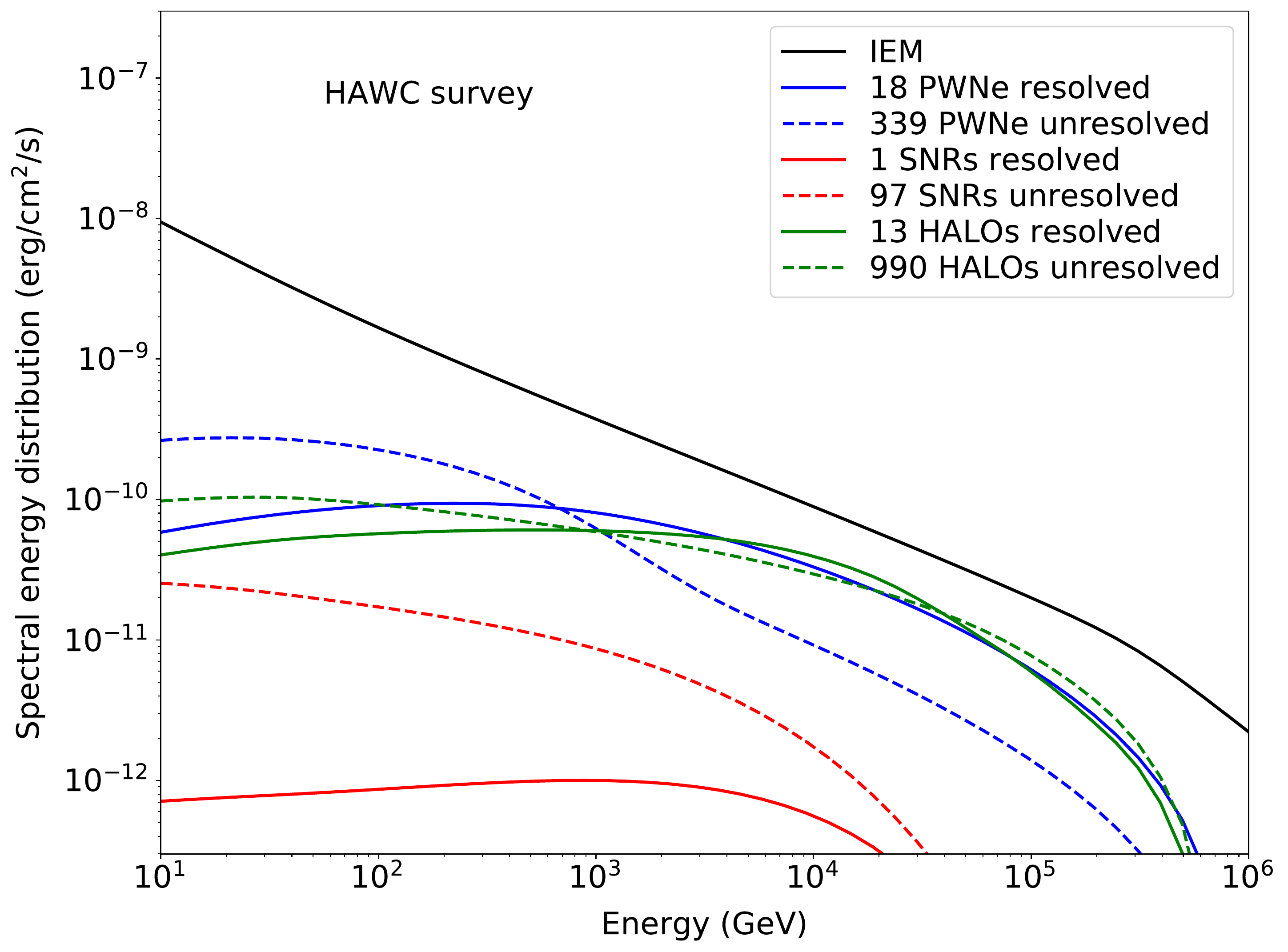}
\includegraphics[width=0.9\columnwidth]{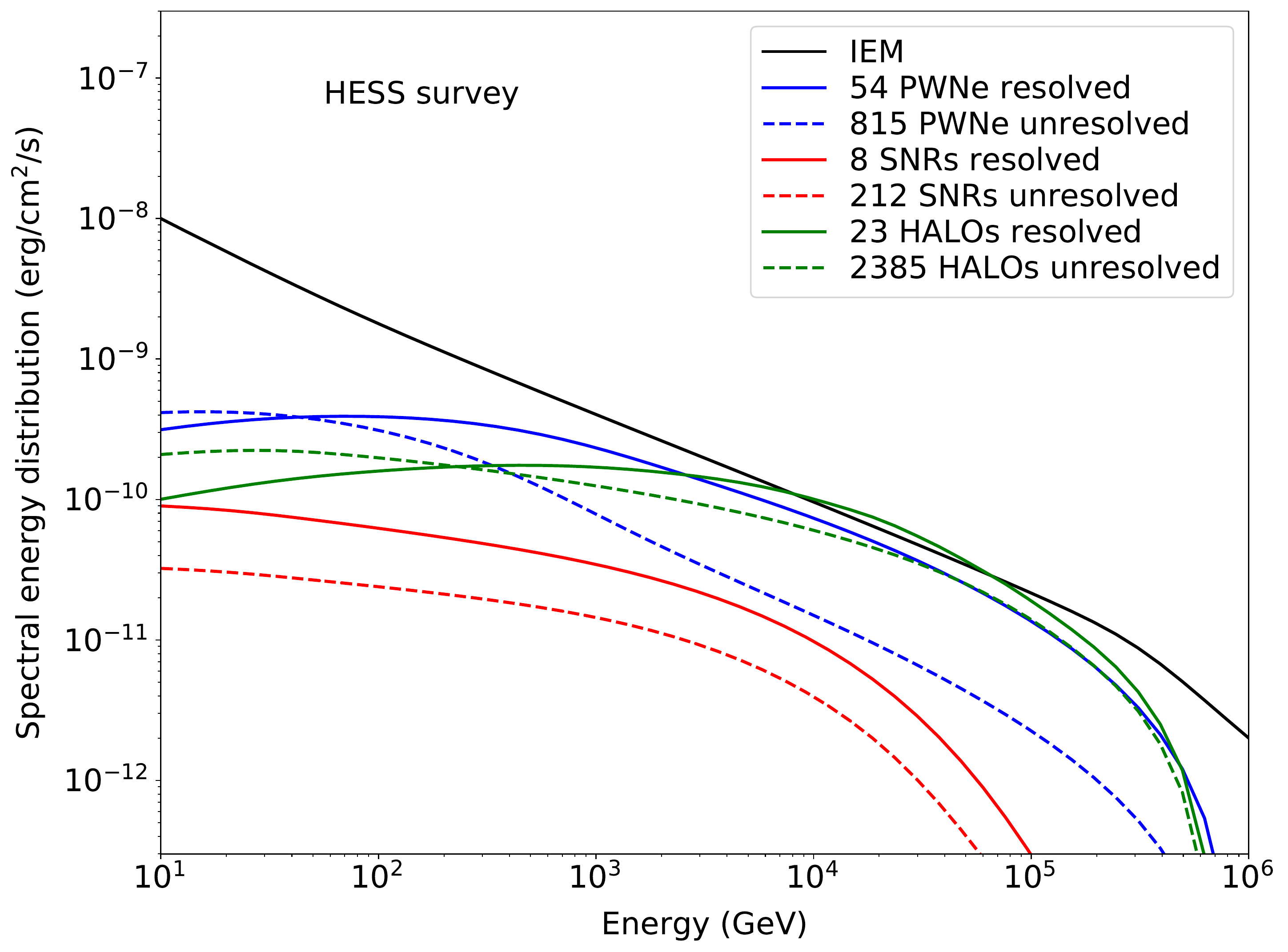}
\includegraphics[width=0.9\columnwidth]{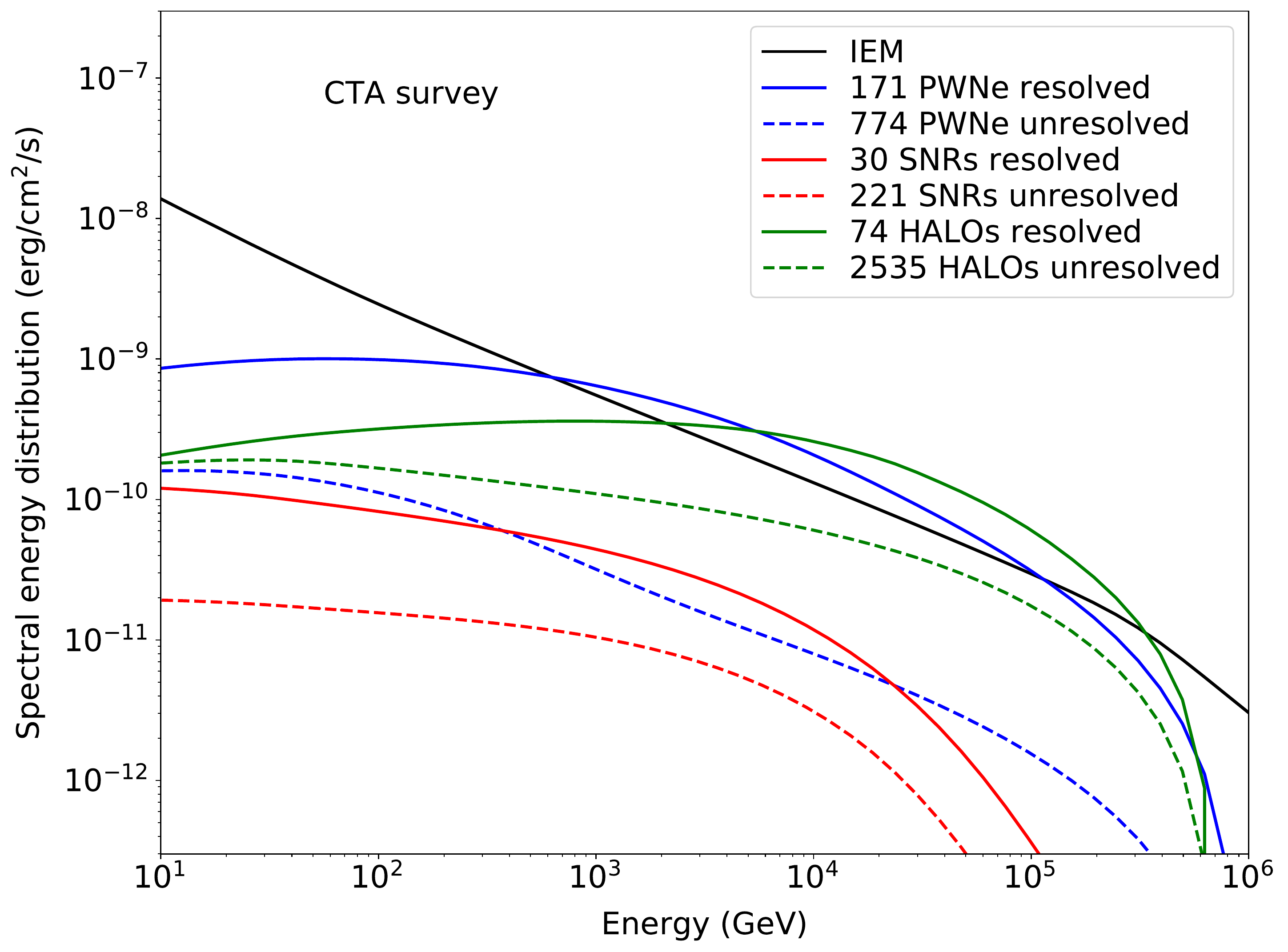}
\caption{Cumulative resolved and unresolved emission for all objects classes in different surveys, compared with the ``Base Max'' interstellar emission model from \citet{DeLaTorreLuque:2022} integrated over the survey footprint (in black, labeled ``IEM''). The mismatch in the number of objects detectable in the HAWC survey between the top panel and Table \ref{tab:detsrcs} is due to the restriction of the footprint (see text).}
\label{fig:res:diffunres}
\end{center}
\end{figure}

\begin{figure}[!t]
\begin{center}
\includegraphics[width=0.9\columnwidth]{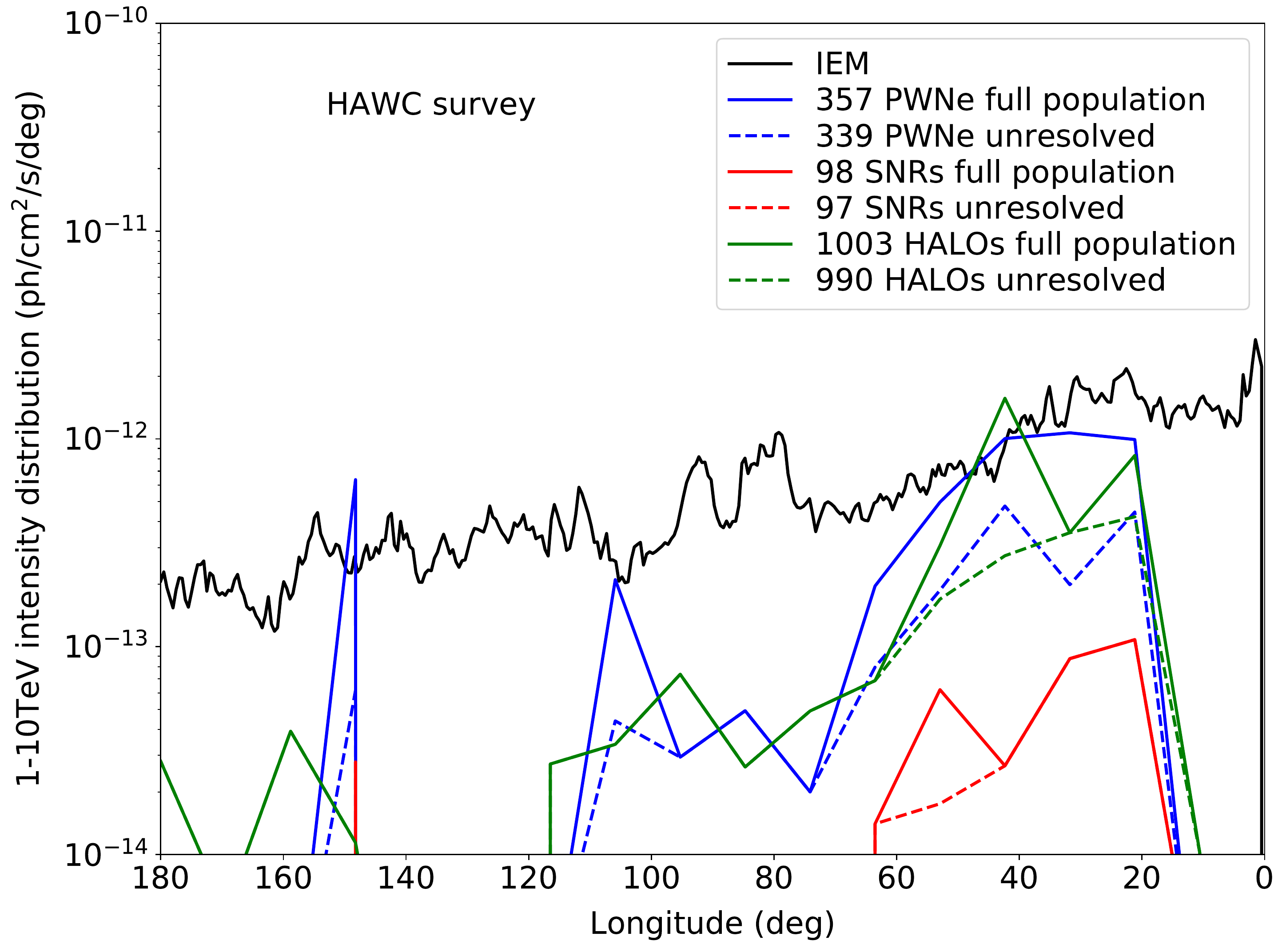}
\includegraphics[width=0.9\columnwidth]{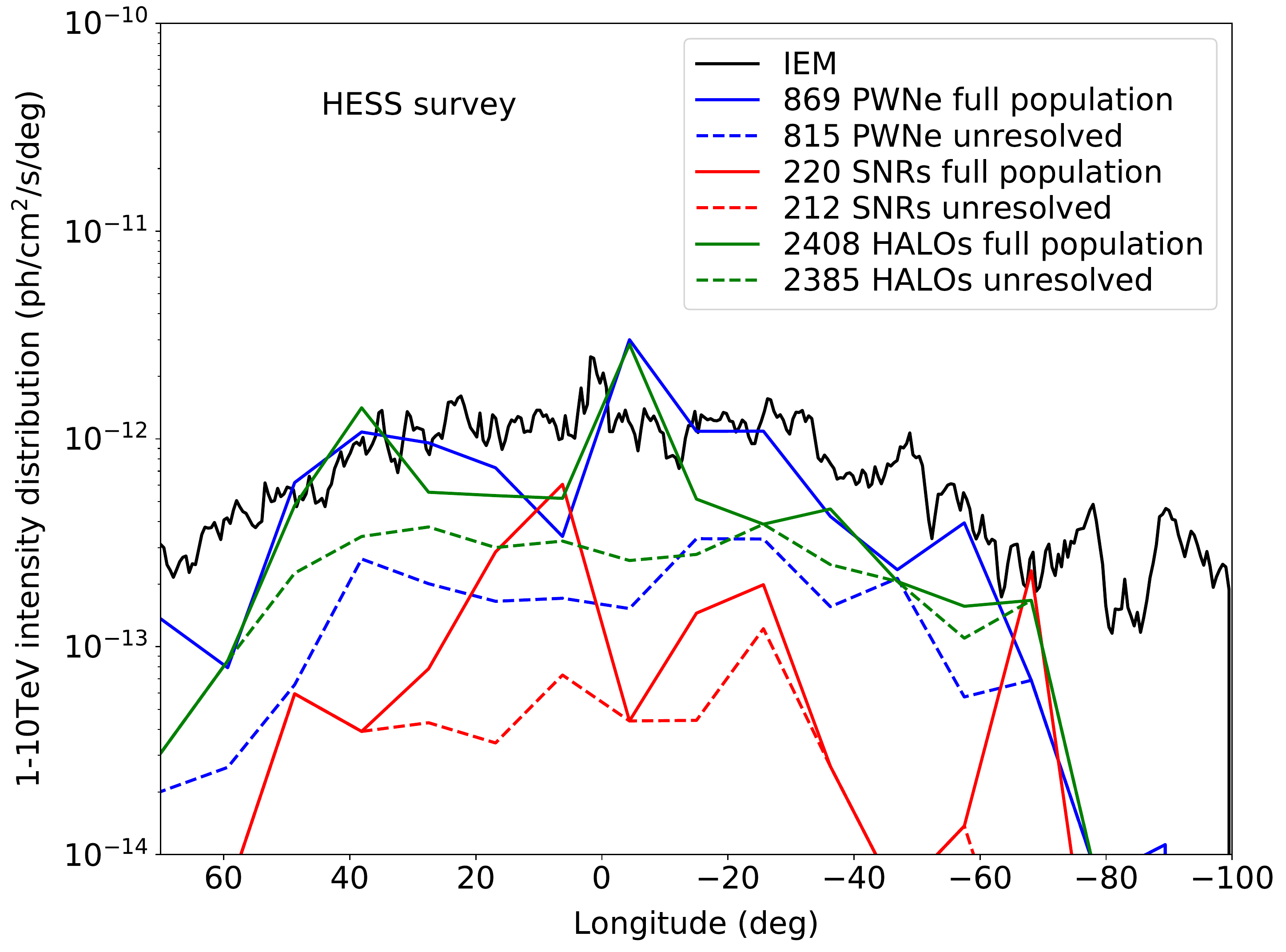}
\includegraphics[width=0.9\columnwidth]{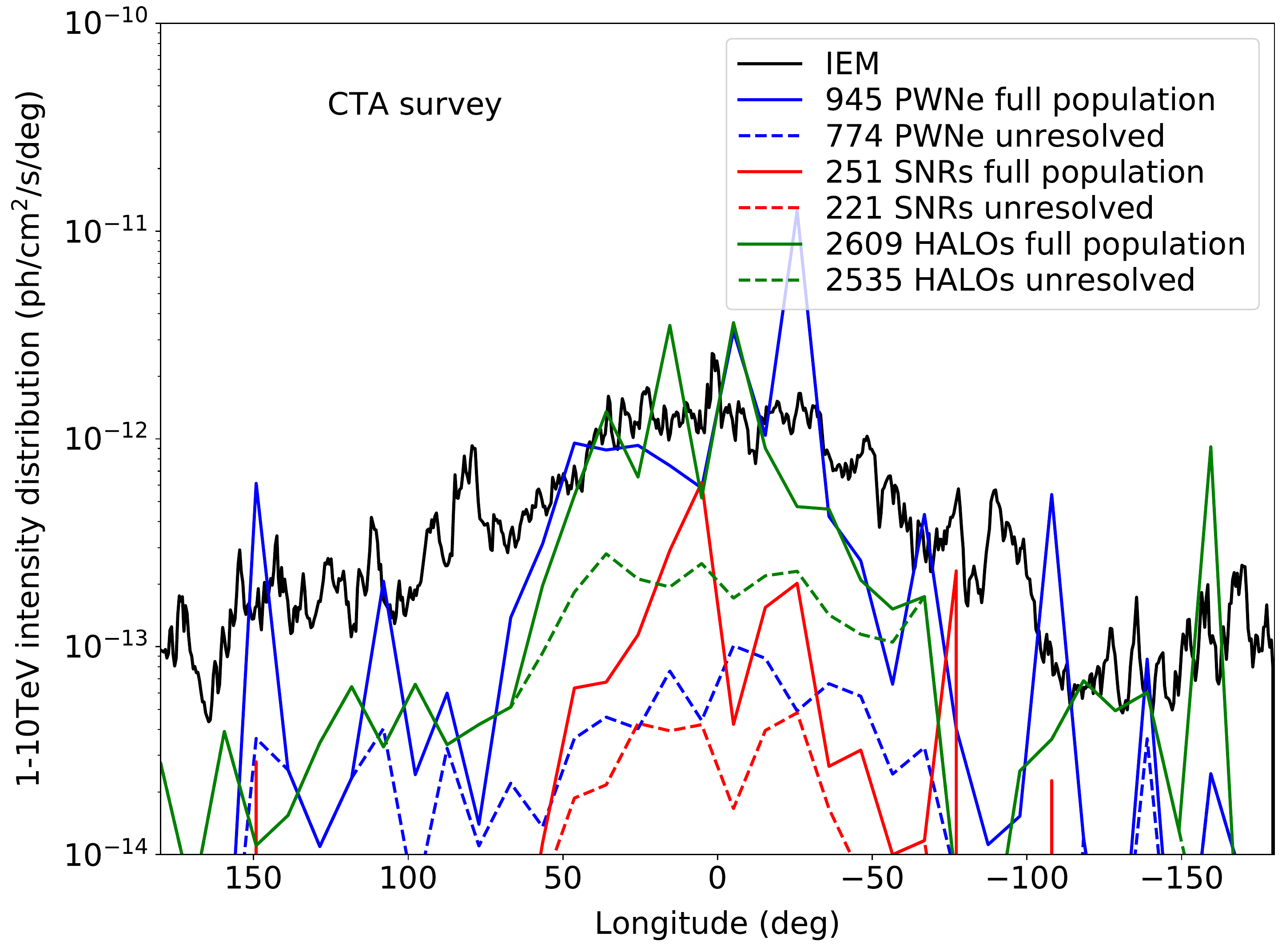}
\caption{Intensity profiles of the full population and unresolved sources in each survey. For comparison, interstellar emission integrated over the footprint of the survey is displayed (for HAWC, the footprint was restricted to [0\deg,180\deg] in longitude and [-6\deg,6\deg] in latitude). For source populations, the emission is summed over 10\deg\ bins in longitude.}
\label{fig:res:profunres}
\end{center}
\end{figure}

We then assess the cumulative emission from all undetectable objects in the surveys considered above, using for detectability the methodology described in Sect. \ref{popres:det}. The results are presented in Fig. \ref{fig:res:diffunres} for the total spectra, and in Fig \ref{fig:res:profunres} for the intensity profiles along the plane. In each panel, the interstellar emission spectrum displayed for comparison was integrated over the footprint of each survey (with a restriction to [0\deg,180\deg] in longitude and [-6\deg,6\deg] in latitude for the HAWC survey).

The HAWC survey probes less than half of the population owing to the location of the instrument in the northern hemisphere. Overall, over the footprint of the survey and in the core energy range of HAWC, the emission from unresolved halos is a factor $2-3$ below that of interstellar radiation, while that of unresolved \gls{pwne} is fainter by about an order of magnitude. The actual distribution of these emission components along the portion of the inner plane that was best surveyed is however rather contrasted, as illustrated in the top panel of Fig. \ref{fig:res:profunres}.

In the H.E.S.S. survey, the emission from unresolved \gls{pwne} is at least a factor $5-6$ fainter than interstellar emission over most of the relevant energy range, while that from unresolved halos is comparable to it, especially above 10\tev. Overall, at core energies for H.E.S.S., the total emission from resolved or unresolved pulsar-powered sources and interstellar radiation are predicted to be of similar magnitude, while the contribution from \gls{snrs} is subdominant. 

The CTA survey widens the gap between the total emission from resolved and unresolved sources, especially in the case of \gls{pwne} where the difference reaches more than an order of magnitude over most of the relevant energy range. Eventually, the CTA survey succeeds in pushing the emission from unresolved \gls{pwne} by a factor 20 or more below the level of interstellar radiation, leaving only unresolved halos as a comparable contribution at energies above 10\tev.

We emphasize that the above statements come with the caveat that detectability was assessed from the simple criterion that the flux be above the survey sensitivity. In reality, source confusion and complicated emission morphologies will most likely tend to lower the detectable fraction and enhance the unresolved contribution. Even more important in the case of halos is the fact that their unresolved emission is estimated based on the limit assumption that all middle-aged pulsars do develop a halo. If only a small fraction of them do so, in the range $5-10$\% as suggested in \citet{Martin:2022}, the above results need to be rescaled accordingly. In such a case, the emission from halos as a whole would be subdominant compared to interstellar radiation (see Fig. \ref{fig:res:diffism}) and their unresolved emission would be a minor component in all surveys (see Fig. \ref{fig:res:diffunres}). Both statements are all the more true that our reference model for interstellar radiation is a minimal prediction; improved models yielding better fits to gamma-ray observations result in emission levels higher by 20-30\% in the $1-10$\tev range \citep{DeLaTorreLuque:2022}.

% Local positron flux from the halo population
\subsection{Local positron flux from the halo population}
\label{popres:posi}

\begin{figure}[!t]
\begin{center}
\includegraphics[width=0.9\columnwidth]{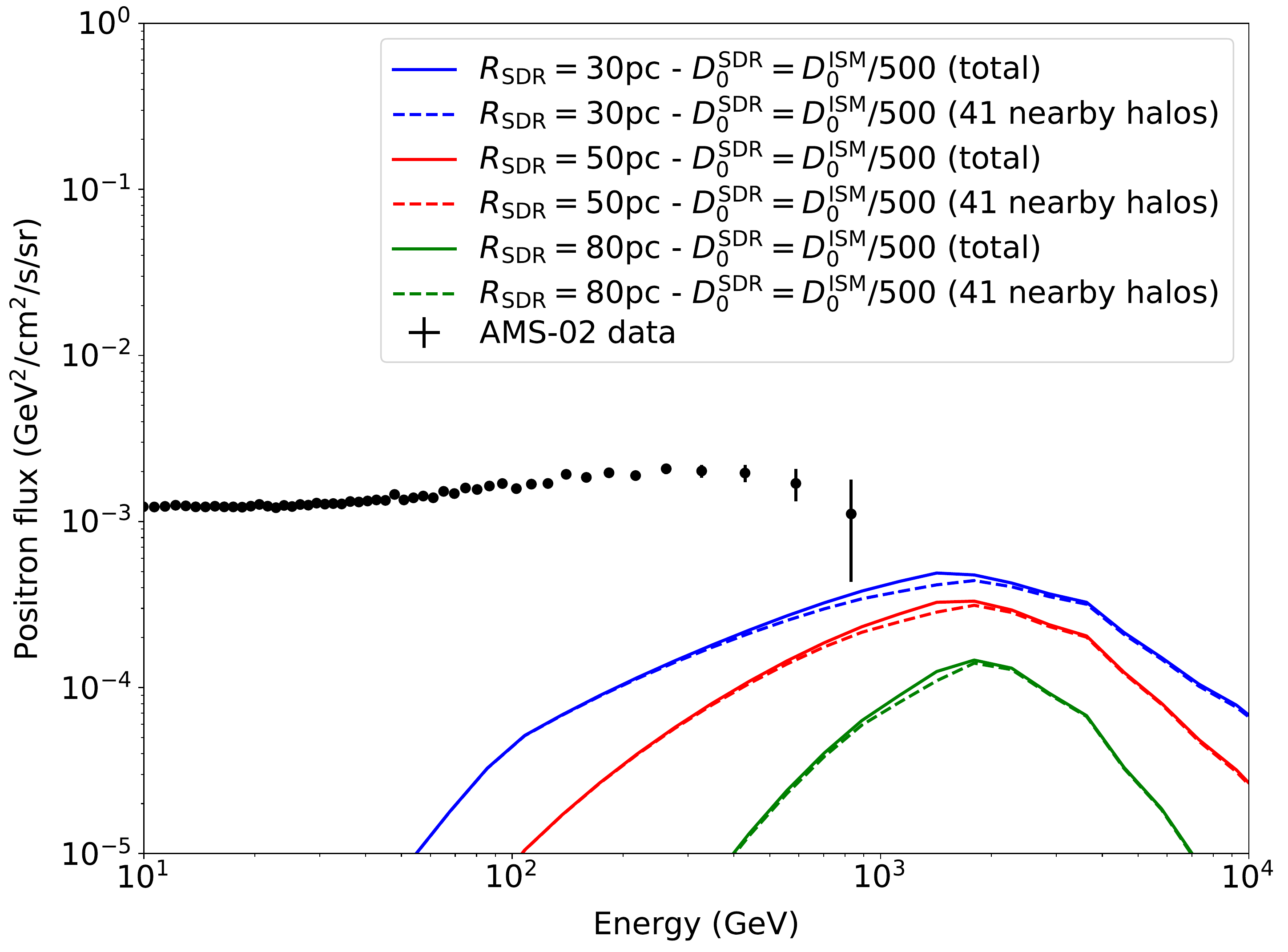}
\caption{Local positron flux from the mock population of pulsar halos, for three model setups with suppressed diffusion region of sizes 30, 50, or 80\,pc and diffusion suppression by a factor 500. The contribution of 41 nearby halos located within 2\,kpc is shown with dashed lines. The predicted fluxes are compared to the AMS-02 measurement.}
\label{fig:res:posflux}
\end{center}
\end{figure}

Figure \ref{fig:res:posflux} shows the local flux of positrons from all mock halos, for the three suppressed diffusion region sizes 30, 50, or 80\,pc and diffusion suppression by a factor 500, compared to the AMS-02 measurement from \citet{Aguilar:2019a}. As illustrated in the plot, most of the local positron flux is contributed by three dozens of nearby objects within 2\,kpc and the corresponding spectrum peaks at an energy of about 2\tev. Particles of higher energies are limited in range owing to strong energy losses, while particle of lower energies cannot diffuse efficiently up to Earth within the age of their parent pulsars (set in our population model to a maximum of 400\,kyr). An enhanced confinement enforced by a larger suppressed diffusion region results in a smaller positron flux and a more peaked spectrum. 

The total predicted positron flux from the whole population is consistent with existing measurements below 1\tev for all considered sizes of the suppressed diffusion region. Such a result complements the work presented in \citet{Martin:2022}, where the contribution of putative nearby halos within 1\,kpc was studied. Indeed, as illustrated in Fig. \ref{fig:res:dists}, the spatial distribution adopted for pulsars yields a deficit of objects within $\lesssim1$\,kpc, as a result of the local arm not being included in the model. 

In \citet{Martin:2022}, it was demonstrated that known middle-aged pulsars within 1\,kpc alone would saturate the measured positron flux if they were to all develop halos with intermediate sizes $30-80$\,pc, diffusion suppression levels like those around J0633+1746 or B0656+14, and injection efficiencies significantly smaller than those inferred for the canonical halos in J0633+1746 and B0656+14, and more generally with the values typical of younger \gls{pwne}. Conversely, if positrons from nearby pulsars besides J0633+1746 or B0656+14 are released in the \gls{ism} without any confinement around the pulsars, the total positron flux fits into the observed spectrum for similar injection efficiencies of a few tens of percent for all pulsars, from kyr-old objects powering \gls{pwne} to 100 kyr-old objects like J0633+1746 and B0656+14. This led to the suggestion that pulsar halos may be a rare phenomenon, with an occurrence rate as low as $5-10$\% among middle-aged pulsars, although the evidence supporting that depends on the exact properties of the local pulsar population and on the uncertain physics driving the formation and evolution of halos. 

If halos are indeed rather rare, \citet{Martin:2022} show that the local positron flux in the $0.1-1$\tev range can be attributed to $2-3$ dozens nearby middle-aged pulsars within 1\,kpc, releasing pairs into the \gls{ism} without confinement at the source. Particles escaping from the halo around J0633+1746 would contribute over part or most of the range, depending on the exact properties of the halo, while B0656+14 would have a maximum contribution at 10\tev. The population synthesis presented here complements that conclusion by showing that the contribution from all pulsars at $1-2$\,kpc distances would be negligible in that picture. 

% Conclusions
\section{Conclusions}
\label{conclu}

We presented a modeling of the populations of \gls{snrs}, \gls{pwne}, and pulsar halos in the Milky Way, and assessed their contribution to the \gls{vhe} emission of the Galaxy. For pulsar halos, we assumed by default that they develop around all middle-aged pulsars after pulsar exit from the nebula. We considered three possible extents for the suppressed diffusion region, from 30 to 80\,pc, and two diffusion suppression levels, by a factor of 500 and 50, representative of the values inferred for the halos around PSR J0633+1746 and B0656+14, respectively. The realization of the mock population we worked on features about 260 \gls{snrs}, 950 \gls{pwne}, and 2600 halos.

Focusing on pulsar-powered objects, expected to be the dominant emitters in the \gls{vhe} range, the mock population seems to account satisfactorily for the properties of currently known objects. The TeV flux distribution is well reproduced from the highest fluxes down to $5-10$\% Crab. In this range, the predicted \gls{pwne} population does not saturate the flux distribution of all known Galactic objects, thus leaving room for another class of emitters as likely counterparts to the currently unidentified sources. Pulsar halos are shown to be a viable solution.

Assessing the detectability in existing surveys with H.E.S.S. and HAWC or the planned survey of the Galactic plane with \gls{cta} yields the following prospects: $\sim80$ sources in the H.ES.S. survey, including $\sim20$ halos and $\sim50$ \gls{pwne}; $\sim40$ sources in the 1523-day HAWC survey, which slightly undershoots the actual number, with even proportions of \gls{pwne} and halos; $\sim300$ sources in the planned CTA survey, more than half of which are \gls{pwne} and $50-100$ are halos. If the CTA survey reaches a sensitivity of 1\,mCrab, prospects inflate up to a total of $400-500$ objects including $100-200$ halos.

The large number of individually unresolved \gls{pwne} and halos in existing surveys feeds a significant diffuse emission compared to interstellar radiation powered by the large-scale population of \gls{crs}. In the H.E.S.S. survey of the Galactic plane, the emission from unresolved halos is comparable to interstellar radiation above 10\tev, while that from \gls{pwne} is at least a factor $5-6$ fainter above 100\gev. The planned CTA survey may help us reduce the emission from unresolved \gls{pwne} by a factor 20 or more below the level of interstellar radiation, leaving only unresolved halos as a comparable contribution at energies above 10\tev. This underlines the importance of nailing down the commonness of the phenomenon in the Galaxy. If only a small fraction of middle-aged pulsars do develop a halo, $5-10$\% of them as suggested in another work from the local positron flux constraint \citep{Martin:2022}, the emission from halos as a whole becomes subdominant compared to interstellar radiation and their unresolved emission would be a minor component in all surveys.

If pulsar halos are rare, the total number of currently known \gls{vhe} sources, including unidentified ones, cannot be explained within our model. The mock \gls{pwne} population cannot account for the number and flux distribution of established or candidate \gls{pwne} and unidentified Galactic sources taken together. This points either to the need for a better modeling of \gls{pwne}, or to the possibility that another class of emitters not modeled in this work makes up the bulk of currently unidentified sources. 

Interestingly, an alternative \gls{pwne} population synthesis, based on a more complete framework including the effect of reverberation in the dynamical evolution of the nebula, yields the same prediction: fewer synthetic \gls{pwne} than the total number of observed sources at intermediate fluxes \citep{Fiori:2022}. Continued efforts to model \gls{pwne} along their full evolutionary path and assess their contribution to the \gls{vhe} sky seem warranted. At the very least, the present work shows that middle-aged pulsars bear some potential to account for a significant fraction of currently unidentified TeV sources, but maybe not in the form of halos as they are described today.

\begin{acknowledgements}
The authors acknowledge financial support by CNES for the exploitation of {\it Fermi}-LAT observations, and by ANR for support to the GAMALO project under reference ANR-19-CE31-0014. We thank Pierre Cristofari for sharing his code to model supernova remnants, and Katrin Egberts and Barbara Olmi for useful comments on the work. This work has made use of the SIMBAD database, operated at CDS, Strasbourg, France, and of NASA's Astrophysics Data System Bibliographic Services. The preparation of the figures has made use of the following open-access software tools: Astropy \citep{Astropy:2013}, Matplotlib \citep{Hunter:2007}, NumPy \citep{VanDerWalt:2011}, and SciPy \citep{Virtanen:2020}.
\end{acknowledgements}

\bibliographystyle{aa}
\bibliography{Biblio/Halo.bib,Biblio/ISM.bib,Biblio/ISRF.bib,Biblio/Pulsars.bib,Biblio/DataAnalysis.bib,Biblio/Physics.bib,Biblio/Fermi.bib,Biblio/CosmicRayEscape.bib,Biblio/CosmicRayAcceleration.bib,Biblio/CosmicRayTransport.bib,Biblio/CosmicRayMeasurements.bib,Biblio/GalacticDiffuseEmission.bib,Biblio/SNobservations.bib,Biblio/SNRobservations.bib,Biblio/LMC.bib,Biblio/CTA.bib}

\end{document}